\documentclass[reprint,prb,amsmath,amssymb,aps,superscriptaddress]{revtex4-1}

\pdfoutput=1
\usepackage{graphicx}
\usepackage{dcolumn}
\usepackage{bm}
\usepackage{longtable}
\usepackage[T1]{fontenc} 
\usepackage{lipsum} 
\usepackage{amsthm}
\usepackage{amsfonts}
\usepackage{mathtools}
\usepackage{booktabs}
\usepackage{amsmath}
\usepackage{blindtext}
\usepackage{setspace}
\usepackage{makeidx}
\usepackage{rotating}
\usepackage{tikz}
\usepackage{gensymb}
\usepackage{float}

\begin{document}

\title{Electric field and strain induced Rashba effect in hybrid halide perovskites}
\author{Linn Leppert}
\affiliation{Molecular Foundry, Lawrence Berkeley National Laboratory, Berkeley, California 94720, USA}
\affiliation{Department of Physics, University of California Berkeley, Berkeley, California 94720, USA}
\author{Sebastian E. Reyes-Lillo}
\affiliation{Molecular Foundry, Lawrence Berkeley National Laboratory, Berkeley, California 94720, USA}
\affiliation{Department of Physics, University of California Berkeley, Berkeley, California 94720, USA}
\author{Jeffrey B. Neaton}
\affiliation{Molecular Foundry, Lawrence Berkeley National Laboratory, Berkeley, California 94720, USA}
\affiliation{Department of Physics, University of California Berkeley, Berkeley, California 94720, USA}
\affiliation{Kavli Energy NanoScience Institute at Berkeley, Berkeley, California 94720, USA}
\email{jbneaton@lbl.gov}

\begin{abstract}
Using first principles density functional theory calculations, we show how Rashba-type energy band splitting in the hybrid organic-inorganic halide perovskites APbX$_3$ (A=CH$_3$NH$_3^+$, CH(NH$_2$)$_2^+$, Cs$^+$ and X=I, Br) can be tuned and enhanced with electric fields and
anisotropic strain. In particular, we demonstrate that the magnitude of the Rashba splitting of tetragonal (CH$_3$NH$_3$)PbI$_3$ grows with increasing macroscopic alignment of the organic cations and electric polarization, indicating appreciable tunability with experimentally-feasible applied fields, even at room temperature. Further, we quantify the degree to which this effect can be tuned via chemical substitution at the A and X sites, which alters amplitudes of different polar distortion patterns of the inorganic PbX$_3$ cage that directly impact Rashba splitting. In addition, we predict that polar phases of CsPbI$_3$ and (CH$_3$NH$_3$)PbI$_3$ with $R3c$ symmetry possessing considerable Rashba splitting might be accessible at room temperature via anisotropic strain induced by epitaxy, even in the absence of electric fields.
\end{abstract}

\maketitle
Organic-inorganic hybrid halide perovskites have received considerable attention in the photovoltaic community owing to their high power conversion efficiencies achieved within only a few years of device research \cite{Stranks2015b}. First principles calculations have played an important role in the development of these materials, and in particular in the prediction of a range of novel electronic and structural phenomena, such as ferroelectric polarization \citep{Stroppa2014b, Stroppa2015a, Grote2014a}, Rashba and Dresselhaus energy band splitting \cite{Brivio2014,Even2014b, Robles2015a, Zheng2015, Etienne2016a} and non-trivial topological phases \cite{Liu2016}. The Rashba effect is an energy level splitting originating from spin-orbit interactions in systems with broken inversion symmetry, as originally described by Rashba and Dresselhaus in noncentrosymmetric zinc blende \cite{Dresselhaus1955} and wurtzite \cite{Rashba1960} semiconductors, respectively. It has been confirmed in a wide variety of materials with either interfacial or bulk inversion symmetry breaking \cite{Manchon2015}. For example, a "giant" bulk Rashba splitting, characterized by a Rashba coefficient of 3.8\,eV{\AA} has been found in the layered semiconductor BiTeI \cite{Ishizaka2011}. In ferroelectrics, i.e., systems with a spontaneous macroscopic polarization which is switchable by an applied electric field, the inversion symmetry-breaking potential gradient originates from the polarization, allowing for the Rashba splitting to be controlled and switched by an external electric field. 

Recently, a significant Rashba effect of $\sim$2--3\,eV{\AA} has been predicted for the hybrid halide perovskite methylammonium lead iodide, (CH$_3$NH$_3$)PbI$_3$ (MAPbI$_3$), using first-principles calculations \cite{Kim2014f, Amat2014, Robles2015a, Zheng2015}, raising hopes that the compound might find application as a ferroelectric Rashba material in spintronic devices. These calculations, however, rely on structural models for MAPbI$_3$ that assume polar distortions and, with a few exceptions \cite{Etienne2016a}, do not account for finite temperature effects. In light of ample experimental evidence \cite{Beilsten-Edmands2015, Fan2015, Govinda2016, Hoque2016a} and entropic arguments \cite{Filippetti2015a}, which have refuted earlier reports of a ferroelectric or on-average polar phase of MAPbI$_3$ and related materials \cite{Gesi1997, Stoumpos2013a, Kutes2014a}, the question remains whether such large Rashba splitting is globally experimentally accessible or might be tunable at room temperature.

In this Letter, we predict with first principles calculations that a Rashba effect can be observed in MAPbI$_3$ at room temperature with an applied electric field, and quantify how its magnitude is affected by the macroscopic electronic polarization. The magnitude of the energy band splitting depends on the degree of alignment of the organic moieties, which can be achieved via polar distortions that couple directly to electric fields. We further demonstrate that the displacement patterns, and consequently the magnitude of the Rashba splitting at the valence and conduction band edges, can be controlled by chemical substitution at the A site, e.g., by an organic molecule with distinct geometry such as formamidinium (FA), CH(NH$_2$)$_2^+$. Finally, we investigate the existence of novel polar phases of CsPbI$_3$ and MAPbI$_3$ and predict that epitaxial strain can lead to an $R3c$ polar phase with significant Rashba splitting at room temperature.

MAPbI$_3$ is known to undergo two phase transitions with decreasing temperature: from cubic ($Pm\bar{3}m$) to tetragonal ($I4/mcm$) at $T=327$\,K, and from tetragonal to orthorhombic ($Pnma$) at $T=162$\,K \cite{Poglitsch1987}. All three phases are centrosymmetric and feature corner-sharing PbI$_6$ octahedra. Neutron scattering experiments have demonstrated that the MA molecules exhibit four-fold rotational symmetry about their C-N axis and three-fold rotational symmetry around the C-N axis in the $Pm\bar{3}m$ and $I4/mcm$ phases \cite{Chen2015q}. At room temperature and higher, the dynamics of these rotations are believed to be so facile that MA can rotate quasi-randomly \cite{Mattoni2015c}. Upon decreasing the temperature, the rotational motion is dominated by the molecules' high-symmetry orientations, accompanied by a monotonic increase of the rotational angle of the PbI$_6$ octahedra, which is the order parameter of the $Pm\bar{3}m$ to $I4/mcm$ phase transition \cite{Kawamura2002}. Finally, at $T\lesssim 162$\,K, rotations about the C-N axis freeze out and the $Pnma$ phase is realized.

The rotational dynamics of the organic cation are generally not taken into account in static density functional theory calculations of the $Pm\bar{3}m$ and $I4/mcm$ phases of MAPbI$_3$ and other hybrid perovskites. Instead, one or several fixed orientations of the molecules are chosen, and single-point or average properties are computed and reported. Since the interaction of MA with the inorganic PbI$_3$ cage sensitively depends on the orientation of the C-N axis \cite{Lee2015}, the spread of predicted band edge structures and band gaps can largely be attributed to differences in the assumed molecular orientation \cite{Quarti2014a}. Furthermore, structural relaxations for fixed MA orientations can lead to spurious distortions of the inorganic cage, which are largely suppressed at finite temperatures due to the thermal motion of the molecules.

Our density functional theory (DFT) calculations are performed within the Perdew-Burke-Ernzerhof (PBE) generalized gradient approximation and the projector augmented wave formalism (PAW) \cite{Blochl1994, Kresse1999} as implemented in VASP \cite{Kresse1993, Kresse1996}. We treat 9 valence electrons explicitly for Cs (5s$^2$5p$^6$6s$^1$), 14 for Pb (6s$^2$5d$^{10}$6p$^2$), 7 for I (5s$^2$5p$^5$), 7 for Br (4s$^2$4p$^5$), 5 for N (2s$^2$2p$^3$) and 4 for C (2s$^2$2p$^2$).
Spin-orbit coupling (SOC) is taken into account self-consistently. Brillouin zone integrations are performed on $6\times 6\times 6$ $\Gamma$-centered $\mathbf{k}$-point meshes using a Gaussian smearing of 0.01\,eV \cite{Elsasser1994} and a plane wave cutoff of 500\,eV such that total energy calculations are converged to within $\sim$10\,meV. It has been noted earlier that in Pb-based hybrid halide perovkites, PBE and other gradient-corrected density functionals lead to a fortuitous agreement with experimental band gaps as a result of neglecting spin-orbit interactions and many-body effects \cite{Brivio2014}. We have confirmed that the band dispersion and magnitude of the Rashba splitting obtained using PBE+SOC agree well with those obtained using the screened-exchange functional HSE+SOC (see Supporting Information) and hence report PBE results throughout this work.

We perform structural optimizations without SOC, relaxing all ions without imposing symmetry constraints until Hellmann-Feynman forces are less than 0.01\,eV/{\AA}. Taking van der Waals (vdW) interactions into account is important for reaching quantitative agreement with experimental lattice parameters \cite{Egger2014}. In particular, inclusion of vdW interactions reduces the unit cell volume of MAPbI$_3$ by $\sim$5\%, potentially suppressing the polar instability. We use Tkatchenko-Scheffler vdW corrections (PBE-TS) \cite{Tkatchenko2009} to test this effect and find that the Rashba energy band splitting can be up to 30\% lower in both CBM and VBM as compared to the values predicted with a PBE-relaxed structure (see Supporting Information for details on these calculations). In what follows, we report PBE results unless otherwise noted.

\begin{figure}
 \includegraphics[width=\columnwidth]{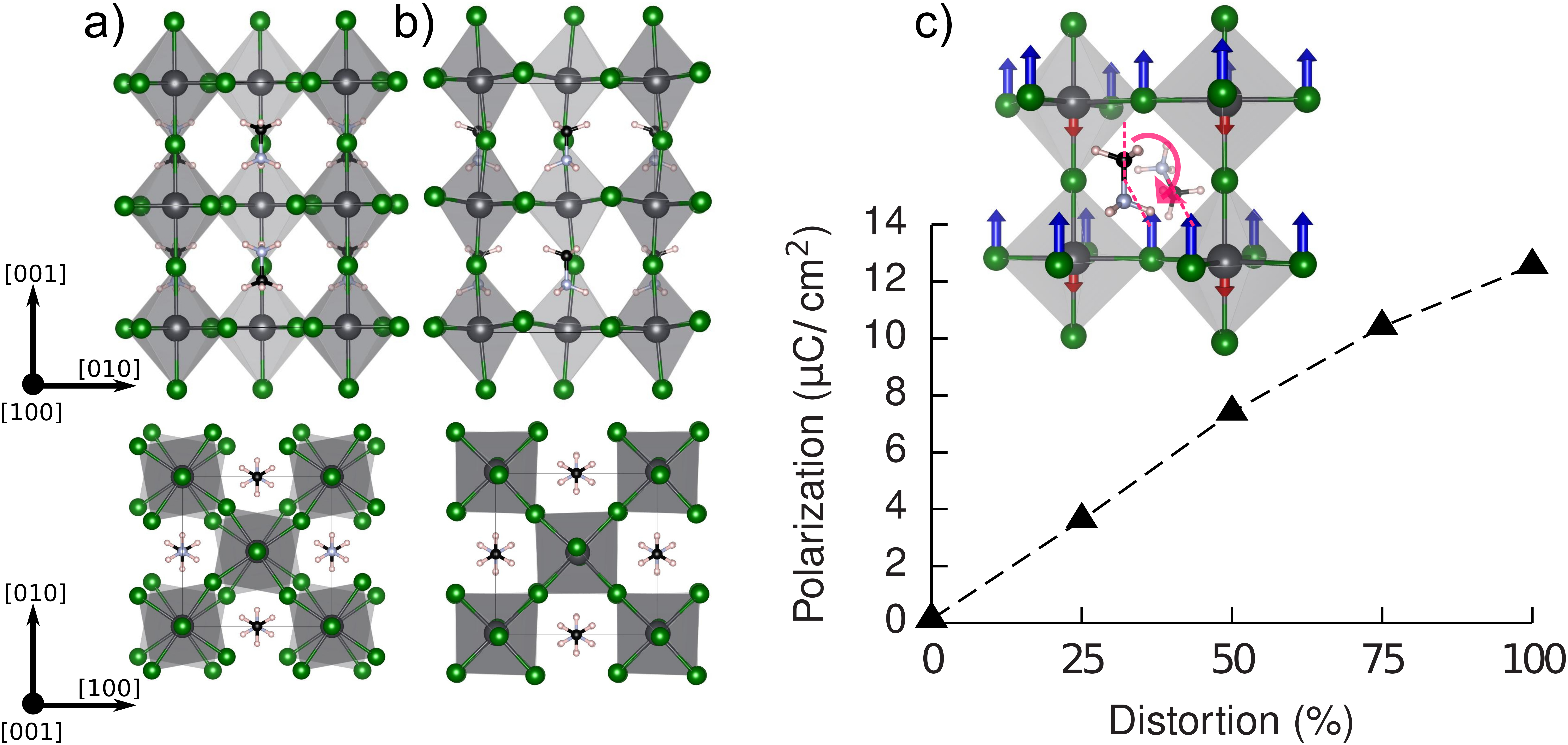}
 \caption{a) Nonpolar $I4/mcm$ reference structure using experimental lattice parameters and atomic positions and MA units aligned anti-parallely. b) Fully relaxed $P4mm$ polar structure with MA units aligned in parallel. c) Polarization as a function of \% distortion along the path from centrosymmetric to polar structure. The inset schematically shows that the polarization is a consequence of polar displacements of Pb (red arrows) and equatorial I (blue arrows) associated with the parallel alignment of MA.}
  \label{fig:path}
\end{figure}
To account for the on-average centrosymmetric structure of MAPbI$_3$ at room temperature and in order not to introduce artifacts related to a fixed orientation of the MA molecules, we start our considerations from the room temperature experimental $I4/mcm$ crystal structure with lattice parameters $a=b=8.876$\,{\AA} and $c=12.553$\,{\AA} \cite{Kawamura2002,Li2015}, using a $\sqrt 2 \times \sqrt 2 \times 2$ unit cell, as shown in Fig.~\ref{fig:path}a). We align the molecules such that they are antiparallel and their net dipole moment is zero, as would be expected on average at room temperature. However, even at finite temperature, MA molecules can be oriented, and polar distortions can be induced, by a sufficiently large electric field. In our calculations, we simulate this situation by performing a structural relaxation starting from the experimental $I4/mcm$ phase with the MAs aligned in parallel. This fully relaxed structure with approximate $P4mm$ symmetry is shown in Fig.~\ref{fig:path}b), and is characterized by vanishing octahedral rotations and polar distortions that are dominated by displacements of the Pb and the equatorial I atoms relative to the $I4/mcm$ reference structure. We construct a structural pathway between the centrosymmetric and the fully polarized structure that consists of a rigid rotation of two of the MA molecules, a decrease of the octahedral rotation amplitude, and an increase in amplitude of the polar distortions.

Using the Berry phase approach within the modern theory of polarization \cite{King-Smith1993}, we calculate the macroscopic polarization of the fully polarized $P4mm$ structure to be 12.6\,$\mu$C/cm$^2$, in very good agreement with previous DFT results \cite{Stroppa2015a}. Fig.~\ref{fig:path}c) quantifies the polarization increase along the structural pathway discussed above from $I4/mcm$ to $P4mm$. In $P4mm$, Pb and apical I atoms are displaced by 0.1\,{\AA} and 0.01\,{\AA} along the [001] direction. The equatorial I atoms experience a displacement of 0.2\,{\AA} along the same axis in the opposite direction (see Supporting Information for a full list of atomic displacements and Born effective charges).
\begin{figure}
 \includegraphics[width=\columnwidth]{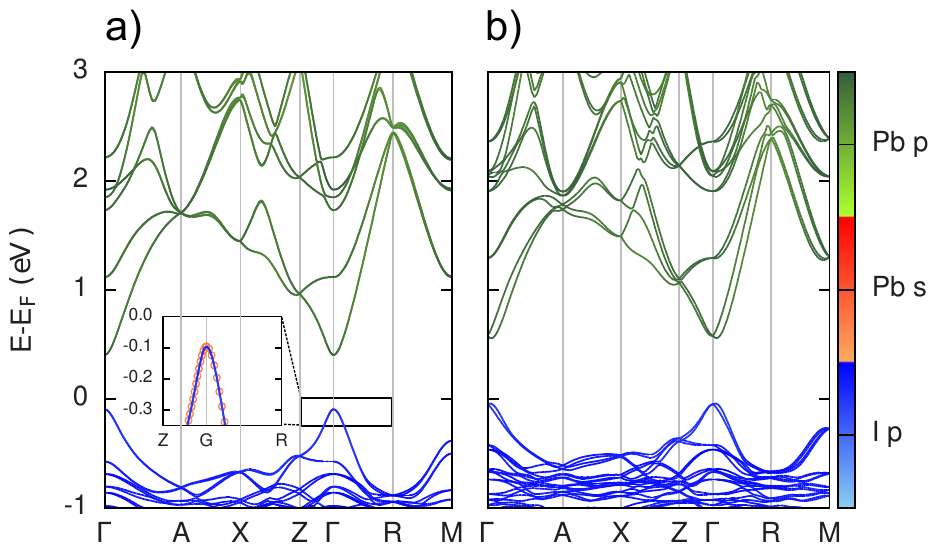}
 \caption{a) Band structure of MAPbI$_3$ in centrosymmetric $I4/mcm$ symmetry, calculated using PBE including SOC. All bands are two-fold degenerate. The colors signify the dominant orbital character of each band. The VBM is predominantly of I p character, but the inset shows that the VBM additionally has Pb s (circles, red color scale) character. b) Band structure of fully polarized MAPbI$_3$. Breaking the inversion symmetry by aligning the molecules, leads to a Rashba splitting of the bands in the directions perpendicular to the distortion (see text).}
  \label{fig:bands}
\end{figure}

We now turn to the evolution of the electronic structure of MAPbI$_3$ along the same structural pathway, focusing on the Rashba splitting of the energy bands in \textbf{k}-space. The band structures of centrosymmetric and polar MAPbI$_3$ are shown in Fig.~\ref{fig:bands}a) and b), respectively. As is well known, the conduction band minimum (CBM) is primarily comprised of Pb $p$-like states, whereas the orbital character of the valence band maximum (VBM) is I $p$ and Pb $s$. Fig.~\ref{fig:bands}b) demonstrates that breaking the inversion symmetry of the structure, lifts the degeneracy of these bands, as has been discussed in previous studies \cite{Brivio2014,Even2014b}. The magnitude of this band splitting is \textbf{k}-dependent, and can be approximately understood from quasi-degenerate perturbation theory \cite{Robles2015a}, where the Rashba Hamiltonian $H_R = \lambda(\mathbf{k}) \cdot \sigma$ with $\lambda(\mathbf k) = \langle \phi_{n \mathbf k} | \frac{\hbar}{4 m^2 c^2}(\nabla \Phi \times (\hbar \mathbf k + \mathbf p)) | \phi_{n \mathbf k} \rangle$ is treated as a perturbation to a zero-order model Hamiltonian without spin-orbit interactions. Here, $\Phi$ is the crystal potential, $\sigma$ the Pauli spin matrices, $\mathbf{p}$ the momentum operator, and $\phi_{n \mathbf k}$ are Kohn-Sham states, $m$ are electron effective masses, and $c$ is the speed of light. In a 3D system with a polar distortion along the [001] direction, $\mathbf k_c=(0,0,\frac{\pi}{c})= \mathbf k_{\parallel}$, where $c$ is the [001] lattice parameter, defines the quantization axis along which the degeneracy of the bands is maintained. This can be seen in Fig.~\ref{fig:bands}b), where the Rashba splitting from $\Gamma(0,0,0)$ to $Z(0,0,\frac{\pi}{c})$ is negligibly small. Along the directions in the plane perpendicular to $\mathbf{k}_{\parallel}$, which in our case can be spanned by the vectors $\mathbf{k}_a=(\frac{\pi}{a},0,0)$ and $\mathbf{k}_b=(0,\frac{\pi}{b},0)$, the Rashba splitting takes on the largest values.  

\begin{figure}
 \includegraphics[width=7cm]{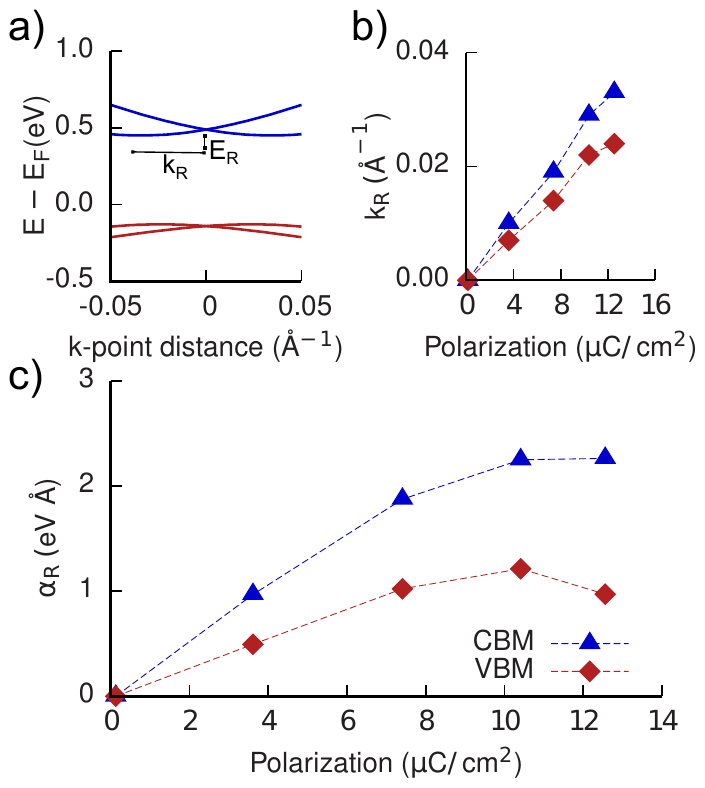}
 \caption{a) The Rashba parameter $\alpha_R$ is defined using the k-space splitting $k_R$ and the energy splitting $E_R$. b) The Rashba splitting $k_R$ increases monotonically for both CBM and VBM. c) $\alpha_R^{\text{CBM}}$ and $\alpha_R^{\text{VBM}}$ as a function of polarization for MAPbI$_3$.}
  \label{fig:Rashba-I}
\end{figure}

We quantify the magnitude of the Rashba effect using the parameter $\alpha_R=2E_R/k_R$, where $k_R$ is the distance in $\mathbf{k}$-space between the crossing point of the spin-split bands and the CBM or VBM, and $E_R$ is the respective energy difference as shown in Fig.~\ref{fig:Rashba-I}a). Since the Rashba splitting is isotropic in the (001) plane, we calculate the band structure from $\Gamma$(0,0,0) to $X$($\frac{\pi}{a}$,0,0) for five structures along the path specified in Fig.~\ref{fig:path} and plot $k_R^{\text{CBM}}$, $k_R^{\text{VBM}}$, $\alpha^{\text{CBM}}_R$ and $\alpha^{\text{VBM}}_R$ as a function of polarization. Fig.~\ref{fig:Rashba-I}b) and c) demonstrate that the Rashba effect in the CBM increases with increasing polarization and reaches a maximum value of $\alpha^{\text{CBM}}_R$=2.3\,eV{\AA}, roughly 60\% of the value reported for BiTeI \cite{Ishizaka2011} and in good agreement with previous reports \cite{Kim2014f,Robles2015a,Zheng2015}. $\alpha^{\text{VBM}}_R$ behaves similarly, and reaches a value about half the size of $\alpha^{\text{CBM}}_R$. The Rashba wavevector $k_R$ increases monotonically for both the CBM and the VBM with polarization. However, the specific choice of structural pathway leads to $E_R^{VBM}$, and consequently $\alpha^{\text{VBM}}_R$, having a maximum at $P \approx 11\,\mu\text{C/cm}^2$. 

We can estimate the electric field, $E_c$, necessary to align the MA molecules and to induce the polar distortions of the $P4mm$ phase to first order as $E_c \approx \frac{\Delta E}{V P}$, where the unit cell volume $V$ and the polarization $P$ are obtained from our first-principles calculations and $\Delta E$ is the energy barrier for the alignment of the MA molecules; for $\Delta E$, we use a recently measured value of 70\,meV \cite{Chen2015q}, which is slightly higher than computed values of between $\sim$20\,meV and $\sim$50\,meV for the room-temperature phase of MAPbI$_3$ \cite{Lee2016c}. The resulting critical field is $E_c \approx 10^6$\,V/cm, a large value, corresponding to 20\,V across a MAPbI$_3$ film of $\sim$200\,nm thickness. However, since the Rashba effect increases with increasing polarization, partial MA alignment at smaller fields should be sufficient to observe Rashba splitting in MAPbI$_3$. Assuming an experimentally feasible bias of 4\,V \cite{Leblebici2016}, that corresponds to a polarization of about 3\,$\mu$C/cm$^2$ following the above considerations, we predict a Rashba effect of $\alpha_R^{\text{CBM}} \approx 1$\,eV{\AA}, a smaller but not insignificant value. 

\begin{figure}
 \includegraphics[width=\columnwidth]{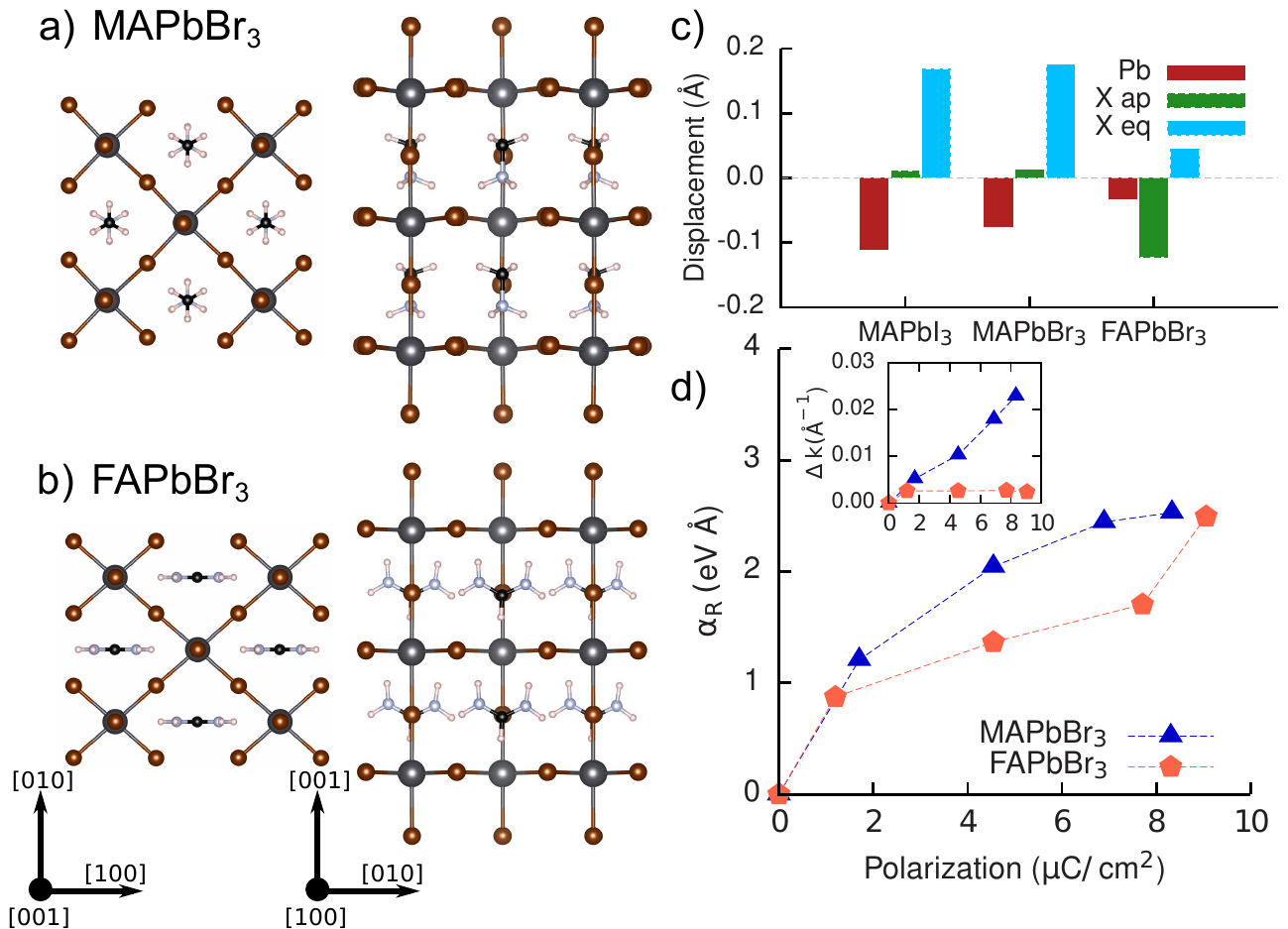}
 \caption{Fully polarized, relaxed structure of a) MAPbBr$_3$ and b) FAPbBr$_3$ c) Displacement of Pb and apical and equatorial halide atoms X=I,Br. d)$\alpha_R^{\text{total}}=\alpha_R^{\text{CBM}}+\alpha_R^{\text{VBM}}$ as a function of polarization for MAPbBr$_3$ and FAPbBr$_3$. The inset shows $\Delta k = k^{\text{CBM}}_R - k^{\text{VBM}}_R$ which is close to zero for FA and increases with increasing polarization for MA.}
  \label{fig:Rashba-Br}
\end{figure}
In Refs.~\citenum{Kim2014f} and \citenum{Zheng2015} it was shown that the spin textures of VBM and CBM can be controlled by realizing different distortion amplitudes and patterns of the perovskite lattice, and in particular by inducing different relative displacements of Pb and apical and equatorial halide atoms. Here we demonstrate that the relative magnitude of the Rashba splitting in VBM and CBM can be controlled in the same way, and that such an effect can be achieved in practice by chemical substitution at the A and X sites. To show this we first replace I by Br and then additionally MA by FA, which has been a commonly used substitute for MA in recent work \cite{Stoumpos2013a,Eperon2014a,Amat2014,Rehman2015a}. Comparing MAPbBr$_3$ and FAPbBr$_3$ is rather straightforward, since, contrary to MAPbI$_3$ and FAPbI$_3$, both are cubic at room temperature. For these two compounds we use structural distortion pathways analogous to those used for MAPbI$_3$, in both cases starting from a centrosymmetric structure with experimental lattice parameters and $Pm\bar{3}m$ symmetry as reference \cite{Rehman2015a,Kulkarni2014a}. The fully relaxed structures with parallely aligned MA and FA molecules are shown in Fig.~\ref{fig:Rashba-Br}a) and b), respectively. Both the displacements of Pb and Br atoms (see Fig.~\ref{fig:Rashba-Br}c)) in MAPbBr$_3$, and the Born effective charges (see Supporting Information), are very similar in magnitude to MAPbI$_3$. The displacement of Pb in MAPbBr$_3$ is about 70\% of that of MAPbI$_3$, which can be attributed to the smaller unit cell volume of the Br compound. The Rashba splitting in CBM and VBM shows a similar trend as a function of polarization, but with a maximum of only $\alpha^{\text{CBM}}_R$=1.9\,eV{\AA} and $\alpha^{\text{total}}_R=\alpha_R^{\text{CBM}}+\alpha_R^{\text{VBM}}$=2.5\,eV{\AA} (Fig.~\ref{fig:Rashba-Br}d)), as the polarization is smaller (8.2\,$\mu$C/cm$^2$) and the SOC in Br is weaker than in I.

Replacing MA with FA changes the picture considerably. Firstly, in the fully polarized structure, the in-plane lattice vectors perpendicular to [001] increase from 8.5\,{\AA} in the nonpolar experimental structure to $a=9.4$\,{\AA} and $b=8.0$\,{\AA} in the fully polarized structure, owing to the two-dimensional geometry of FA. Furthermore, the alignment of FA leads to a small relative Pb atom displacement of -0.03\,{\AA}, whereas the dominant contribution to the distortion along [001] arises from the apical Br atoms, resulting in a polarization of 9.1\,$\mu$C/cm$^2$. The Rashba splitting reaches a maximum of $\alpha^{\text{total}}_R$=2.5\,eV{\AA}, i.e., the same value as in MAPbBr$_3$. Note however, that unlike for MAPbBr$_3$, where the splitting occurs mainly in the CBM, both CBM and VBM exhibit similar amounts of Rashba splitting for FAPbBr$_3$. This is demonstrated in the inset of Fig.~\ref{fig:Rashba-Br}d) which shows the calculated trend in $\Delta k = k^{\text{CBM}}_R - k^{\text{VBM}}_R$ with polarization. $\Delta k$ increases with increasing polarization for MAPbBr$_3$, because the inversion symmetry breaking field along the path changes mainly due to the displacement of the Pb and the equatorial Br atoms. This in turn leads to stronger Rashba splitting for the CBM due to its predominant Pb 6$p$ character. Conversely, in FAPbBr$_3$, the displacement of Pb, and both the equatorial and apical Br atoms, results in inversion symmetry breaking that affects both the CBM and the VBM (predominantly Br 5$p$ and Pb 5$s$ orbital character). 

We now turn to evaluating the possibility of stabilizing a polar phase at room temperature that would allow the observation of Rashba splitting without the need for strong electric fields. In what follows, we investigate low-energy polar phases of CsPbI$_3$ and MAPbI$_3$. From a computational perspective, replacing MA with Cs  avoids complications related to the molecular orientation and provides an approximate way of assessing the effect of biaxial strain on MAPbI$_3$. Our approach is motivated by the well-studied effects of anisotropic strain due to epitaxial growth on the phase stability in particular ferroelectric phases in traditional oxide perovskites \cite{Rabe2007}. Previous computational studies have considered the effect of hydrostatic pressure and biaxial strain \cite{Grote2015, Liu2016}. Furthermore, the experimental stabilization of the cubic $Pm\bar{3}m$ phase of CsPbI$_3$ at room temperature has been attributed to strain \cite{Eperon2015}. However, no studies thus far have considered polar halide perovskites under biaxial strain.

For MAPbI$_3$, we use PBE-TS and a plane wave cutoff energy of 600\,eV to obtain accurate lattice parameters for the experimentally observed centrosymmetric phases $Pnma$, $I4/mcm$ \cite{Poglitsch1987}, as well as the polar $R3c$ phase (see Supporting Information). Tab.~\ref{tab:energies} lists the energetics for each phase and compares them with the corresponding phases of CsPbI$_3$, for which we additionally consider polar $P4mm$, $Amm2$ and $R3m$ structures, as well as the non-perovskite $Pnma$ room-temperature phase of CsPbI$_3$ (n-$Pnma$). We calculate that $R3c$ is the only energetically relevant polar phase for both compounds, with $P4mm$, $Amm2$, and $R3m$ being only $\sim$5\,meV lower in energy than the cubic reference phase. Interestingly, in the case of MAPbI$_3$, antiparallel alignment of the MA units in the $R3c$ structure suppresses the polar $\Gamma_4^-$ mode and results in $R\bar{3}c$ structural symmetry. The alignment of the molecules is associated with a small energy cost of $\sim$35\,meV.

\begin{table}[htbp]
\caption{Energy gain and estimated equilibrium strain of selected phases of CsPbI$_3$ with respect to the high temperature cubic phase. The room temperature phase of CsPbI$_3$ is a non-perovskite structure with $Pnma$ symmetry, here denoted as n-$Pnma$.}
\begin{center}
\begin{tabular}{c|c|c|c}
Cs: space group & $\Delta E$ (meV/f.u.) & $\sigma_{\text{ab}}$ & $\sigma_{\text{bd}}$ \\ \hline \hline
$R3c$ & 76 & --- & -1.0 \\ 
$I4/mcm$ & 84 & -2.2 & -0.4 \\ 
$Pnma$ & 120 & -1.4 & -1.3 \\ 
n-$Pnma$ & 170 & --- & --- \\ \hline
\multicolumn{1}{l}{MA: space group} & $\Delta E$ (meV/f.u.) & $\sigma_{\text{ab}}$ & $\sigma_{\text{bd}}$ \\ \hline \hline
$R3c$ & 49 & --- & -1.0 \\ 
$I4/mcm$ & 97 & -2.0 & -0.6 \\ 
$Pnma$ & 183 & -2.5 & -1.2 \\ 
\end{tabular}
\end{center}
\label{tab:energies}
\end{table}

In the case of CsPbI$_3$, the polar phases $P4mm$ and $Amm2$ exhibit negligibly small Rashba energy band splitting of less than 0.005\,\AA$^{-1}$. The Rashba splitting of the CBM of $R3m$ and $R3c$ is $k_R^{\text{CBM}}$=0.012\,{\AA}$^{-1}$, significantly larger. In $R3c$-MAPbI$_3$, the size of the Rashba splitting approximately doubles compared with the corresponding CsPbI$_3$ phase, with $k_R^{\text{CBM}}$=0.023\,{\AA}$^{-1}$ and $\alpha_R^{\text{CBM}}$=1.6\,eV{\AA} obtained for the $R3c$ structure, highlighting the crucial role of the MA molecule for large Rashba splitting.

\begin{figure*}[ht]
 \includegraphics[width=13cm]{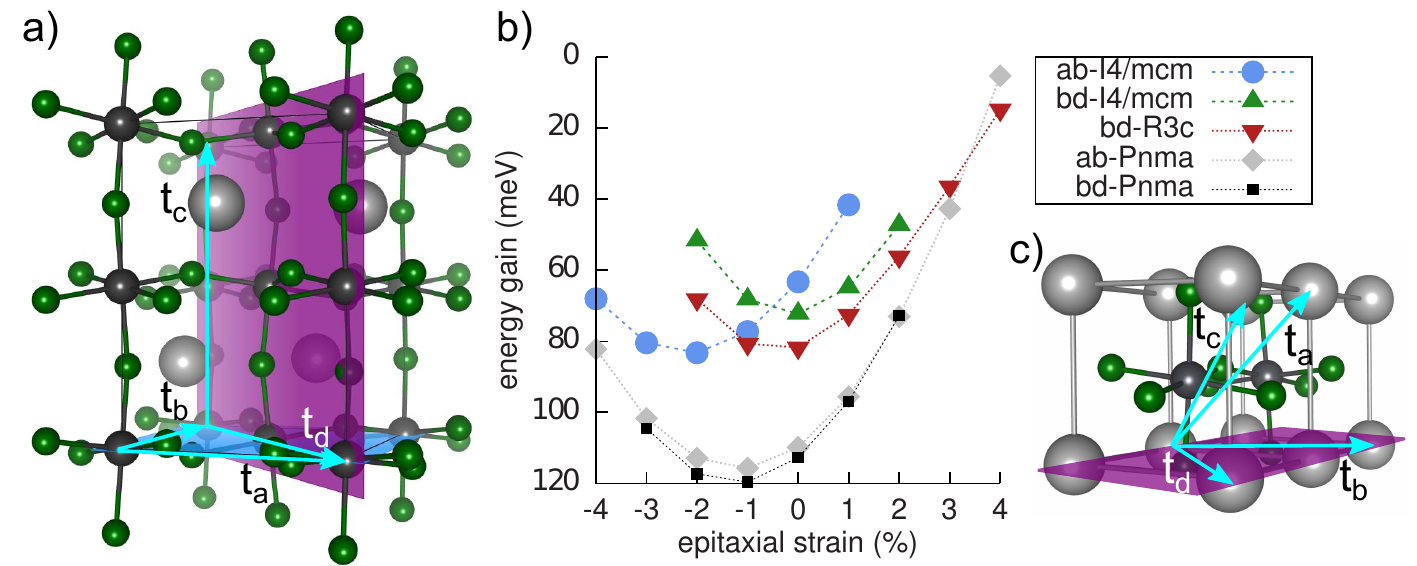}
 \caption{a) Lattice vectors of the $Pnma$ phase. The plane spanned by the vectors $\mathbf{t_a}$ and $\mathbf{t_b}$ (denoted by $j$=ab) is shown in blue, whereas the plane spanned by $\mathbf{t_b}$ and $\mathbf{t_d}$ ($j$=bd) is shown in violet. For the $I4/mcm$ phase the epitaxial matching planes are defined in the same way. b) Epitaxial phase diagram of CsPbI$_3$. c) Lattice vectors of the $R3c$ phase. The matching plane is spanned by $\mathbf{t_b}$ and $\mathbf{t_d}$.}
  \label{epitaxial}
\end{figure*}

To investigate whether the polar $R3c$ phase can be accessed with biaxial strain, we calculate the epitaxial strain diagram of CsPbI$_3$ using "strained-bulk" calculations \cite{Pertsev1998, Dieguez2005}. In both the $I4/mcm$ and the $Pnma$ structure, there are two symmetry-inequivalent epitaxial matching planes, as illustrated in Fig.~\ref{epitaxial}a). The ab-plane (blue) is spanned by the lattice vectors $\mathbf{t}_{\text{a}}$ and $\mathbf{t}_{\text{b}}$, whereas the bd-plane (violet) is spanned by $\mathbf{t}_{\text{b}}$ and $\mathbf{t}_{\text{d}}=\mathbf{t}_{\text{a}}+\mathbf{t}_{\text{b}}$. For $R3c$, where $\mathbf{t}_{\text{d}}=\mathbf{t}_{\text{b}}-\mathbf{t}_{\text{a}}-\mathbf{t}_{\text{c}}$ (Fig.~\ref{epitaxial}c)), $R3c$ is reduced to $Cc$ symmetry under strain. Fig.~\ref{epitaxial}b) shows that the polar $R3c$ phase is stabilized at about 1\% compressive strain, and energetically competes with the $I4/mcm$ phase throughout a range of strains. Above 3\% tensile strain, $R3c$ becomes lower in energy than the $Pnma$ phase due to the suppression of halide octahedral rotations at tensile strain. This suggests that the $R3c$ phase might be realized at room temperature under epitaxial or other forms of large anisotropic strain \cite{Protesescu2015a}.

Since an explicit calculation of the epitaxial strain diagram of MAPbI$_3$ is complicated by the presence of the MA moieties, we follow Ref.~\citenum{Reyes-Lillo2013} to estimate the strain corresponding to the energy minimum of a structural phase as $\sigma_{j}=100 \cdot \frac{1}{2} \sum_i (|\mathbf{t}_i| - |\mathbf{t}_{i0}|)/|\mathbf{t}_{i0}|$. Here $j$ denotes the respective epitaxial matching planes $j=ab$ and $j=bd$. The $\mathbf{t}_{i0}$ refer to the reference lattice vectors constructed from the cubic reference phase. The values reported in Table \ref{tab:energies} for CsPbI$_3$ are close to the respective energy minima in Fig.~\ref{epitaxial}, demonstrating that our method of estimating the equilibrium strain is reliable (see Supporting Information for details). With the exception of the $Pnma$ phase, we find that the $\sigma_j$ of CsPbI$_3$ and MAPbI$_3$ are very similar, suggesting a rather similar energy vs. strain diagram for MAPbI$_3$ and thus the possibility of accessing $R3c$-MAPbI$_3$ with biaxial strain.

In conclusion, we have investigated routes by which Rashba splitting can be observed in room temperature MAPbI$_3$ and related halide perovskites. Due to the rotational freedom of the organic cation, electric fields can break inversion symmetry in $I4/mcm$-MAPbI$_3$ and lead to Rashba splitting. Since the magnitude of the splitting increases with increasing polarization, we expect that the effect will be observable at moderate electric fields that lead to partial MA alignment. We further propose that the band edge characteristics of the splitting can be tuned by inducing different distortive patterns in the Pb-halide cage, and we have considered two examples, substituting MA by FA and anisotropic strain. We predict that under the effect of moderate to high biaxial tensile strain, a polar $R3c$ phase with significant Rashba splitting is accessible for CsPbI$_3$ and MAPbI$_3$, suggesting an alternative experimental route to observe the Rashba effect in halide perovskites.

\begin{acknowledgments}

Work at the Molecular Foundry was supported by the Office of Science, Office of Basic Energy Sciences, of the U.S. Department of Energy, and Laboratory Directed Research and Development Program at the Lawrence Berkeley National Laboratory under Contract No. DE-AC02-05CH11231. LL acknowledges financial support by the Feodor-Lynen program of the Alexander-von-Humboldt foundation.

\end{acknowledgments}

\section*{Supporting Information}
Comparison of Rashba splitting, effective masses and band gaps of $Pm\bar{3}m$-MAPbI$_3$ using PBE, PBE-TS and HSE. Displacements, Born effective charges and atomic coordinates of MAPbI$_3$, MAPbBr$_3$ and FAPbBr$_3$. Further discussion of the epitaxial strain diagram of CsPbI$_3$.


\begin{thebibliography}{51}%
  \makeatletter
\providecommand \@ifxundefined [1]{%
 \@ifx{#1\undefined}
}%
\providecommand \@ifnum [1]{%
 \ifnum #1\expandafter \@firstoftwo
 \else \expandafter \@secondoftwo
 \fi
}%
\providecommand \@ifx [1]{%
 \ifx #1\expandafter \@firstoftwo
 \else \expandafter \@secondoftwo
 \fi
}%
\providecommand \natexlab [1]{#1}%
\providecommand \enquote  [1]{``#1''}%
\providecommand \bibnamefont  [1]{#1}%
\providecommand \bibfnamefont [1]{#1}%
\providecommand \citenamefont [1]{#1}%
\providecommand \href@noop [0]{\@secondoftwo}%
\providecommand \href [0]{\begingroup \@sanitize@url \@href}%
\providecommand \@href[1]{\@@startlink{#1}\@@href}%
\providecommand \@@href[1]{\endgroup#1\@@endlink}%
\providecommand \@sanitize@url [0]{\catcode `\\12\catcode `\$12\catcode
  `\&12\catcode `\#12\catcode `\^12\catcode `\_12\catcode `\%12\relax}%
\providecommand \@@startlink[1]{}%
\providecommand \@@endlink[0]{}%
\providecommand \url  [0]{\begingroup\@sanitize@url \@url }%
\providecommand \@url [1]{\endgroup\@href {#1}{\urlprefix }}%
\providecommand \urlprefix  [0]{URL }%
\providecommand \Eprint [0]{\href }%
\providecommand \doibase [0]{http://dx.doi.org/}%
\providecommand \selectlanguage [0]{\@gobble}%
\providecommand \bibinfo  [0]{\@secondoftwo}%
\providecommand \bibfield  [0]{\@secondoftwo}%
\providecommand \translation [1]{[#1]}%
\providecommand \BibitemOpen [0]{}%
\providecommand \bibitemStop [0]{}%
\providecommand \bibitemNoStop [0]{.\EOS\space}%
\providecommand \EOS [0]{\spacefactor3000\relax}%
\providecommand \BibitemShut  [1]{\csname bibitem#1\endcsname}%
\let\auto@bib@innerbib\@empty
\bibitem [{\citenamefont {Stranks}\ and\ \citenamefont
  {Snaith}(2015)}]{Stranks2015b}%
  \BibitemOpen
  \bibfield  {author} {\bibinfo {author} {\bibfnamefont {S.~D.}\ \bibnamefont
  {Stranks}}\ and\ \bibinfo {author} {\bibfnamefont {H.~J.}\ \bibnamefont
  {Snaith}},\ }\href {http://dx.doi.org/10.1038/nnano.2015.90} {\bibfield
  {journal} {\bibinfo  {journal} {Nat. Nanotechnol.}\ }\textbf {\bibinfo
  {volume} {10}},\ \bibinfo {pages} {391} (\bibinfo {year} {2015})}\BibitemShut
  {NoStop}%
\bibitem [{\citenamefont {Stroppa}\ \emph {et~al.}(2014)\citenamefont
  {Stroppa}, \citenamefont {{Di Sante}}, \citenamefont {Barone}, \citenamefont
  {Bokdam}, \citenamefont {Kresse}, \citenamefont {Franchini}, \citenamefont
  {Whangbo},\ and\ \citenamefont {Picozzi}}]{Stroppa2014b}%
  \BibitemOpen
  \bibfield  {author} {\bibinfo {author} {\bibfnamefont {A.}~\bibnamefont
  {Stroppa}}, \bibinfo {author} {\bibfnamefont {D.}~\bibnamefont {{Di Sante}}},
  \bibinfo {author} {\bibfnamefont {P.}~\bibnamefont {Barone}}, \bibinfo
  {author} {\bibfnamefont {M.}~\bibnamefont {Bokdam}}, \bibinfo {author}
  {\bibfnamefont {G.}~\bibnamefont {Kresse}}, \bibinfo {author} {\bibfnamefont
  {C.}~\bibnamefont {Franchini}}, \bibinfo {author} {\bibfnamefont {M.-H.}\
  \bibnamefont {Whangbo}}, \ and\ \bibinfo {author} {\bibfnamefont
  {S.}~\bibnamefont {Picozzi}},\ }\href {\doibase 10.1038/ncomms6900}
  {\bibfield  {journal} {\bibinfo  {journal} {Nat. Comm.}\ }\textbf {\bibinfo
  {volume} {5}},\ \bibinfo {pages} {5900(1} (\bibinfo {year}
  {2014})}\BibitemShut {NoStop}%
\bibitem [{\citenamefont {Stroppa}\ \emph {et~al.}(2015)\citenamefont
  {Stroppa}, \citenamefont {Quarti}, \citenamefont {{De Angelis}},\ and\
  \citenamefont {Picozzi}}]{Stroppa2015a}%
  \BibitemOpen
  \bibfield  {author} {\bibinfo {author} {\bibfnamefont {A.}~\bibnamefont
  {Stroppa}}, \bibinfo {author} {\bibfnamefont {C.}~\bibnamefont {Quarti}},
  \bibinfo {author} {\bibfnamefont {F.}~\bibnamefont {{De Angelis}}}, \ and\
  \bibinfo {author} {\bibfnamefont {S.}~\bibnamefont {Picozzi}},\ }\href
  {http://dx.doi.org/10.1021/acs.jpclett.5b00542} {\bibfield  {journal}
  {\bibinfo  {journal} {J. Phys. Chem. Lett.}\ }\textbf {\bibinfo {volume}
  {6}},\ \bibinfo {pages} {2223} (\bibinfo {year} {2015})}\BibitemShut
  {NoStop}%
\bibitem [{\citenamefont {Grote}\ \emph {et~al.}(2014)\citenamefont {Grote},
  \citenamefont {Ehrlich},\ and\ \citenamefont {Berger}}]{Grote2014a}%
  \BibitemOpen
  \bibfield  {author} {\bibinfo {author} {\bibfnamefont {C.}~\bibnamefont
  {Grote}}, \bibinfo {author} {\bibfnamefont {B.}~\bibnamefont {Ehrlich}}, \
  and\ \bibinfo {author} {\bibfnamefont {R.~F.}\ \bibnamefont {Berger}},\
  }\href {http://link.aps.org/doi/10.1103/PhysRevB.90.205202} {\bibfield
  {journal} {\bibinfo  {journal} {Phys. Rev. B}\ }\textbf {\bibinfo {volume}
  {90}},\ \bibinfo {pages} {205202} (\bibinfo {year} {2014})}\BibitemShut
  {NoStop}%
\bibitem [{\citenamefont {Brivio}\ \emph {et~al.}(2014)\citenamefont {Brivio},
  \citenamefont {Butler}, \citenamefont {Walsh},\ and\ \citenamefont {van
  Schilfgaarde}}]{Brivio2014}%
  \BibitemOpen
  \bibfield  {author} {\bibinfo {author} {\bibfnamefont {F.}~\bibnamefont
  {Brivio}}, \bibinfo {author} {\bibfnamefont {K.~T.}\ \bibnamefont {Butler}},
  \bibinfo {author} {\bibfnamefont {A.}~\bibnamefont {Walsh}}, \ and\ \bibinfo
  {author} {\bibfnamefont {M.}~\bibnamefont {van Schilfgaarde}},\ }\href
  {\doibase 10.1103/PhysRevB.89.155204} {\bibfield  {journal} {\bibinfo
  {journal} {Phys. Rev. B}\ }\textbf {\bibinfo {volume} {89}},\ \bibinfo
  {pages} {155204} (\bibinfo {year} {2014})}\BibitemShut {NoStop}%
\bibitem [{\citenamefont {Even}\ \emph {et~al.}(2014)\citenamefont {Even},
  \citenamefont {Pedesseau}, \citenamefont {Jancu},\ and\ \citenamefont
  {Katan}}]{Even2014b}%
  \BibitemOpen
  \bibfield  {author} {\bibinfo {author} {\bibfnamefont {J.}~\bibnamefont
  {Even}}, \bibinfo {author} {\bibfnamefont {L.}~\bibnamefont {Pedesseau}},
  \bibinfo {author} {\bibfnamefont {J.~M.}\ \bibnamefont {Jancu}}, \ and\
  \bibinfo {author} {\bibfnamefont {C.}~\bibnamefont {Katan}},\ }\href
  {\doibase 10.1002/pssr.201308183} {\bibfield  {journal} {\bibinfo  {journal}
  {Phys. Stat. Sol. RRL}\ }\textbf {\bibinfo {volume} {8}},\ \bibinfo {pages}
  {31} (\bibinfo {year} {2014})}\BibitemShut {NoStop}%
\bibitem [{\citenamefont {Robles}\ \emph {et~al.}(2015)\citenamefont {Robles},
  \citenamefont {Katan}, \citenamefont {Sapori}, \citenamefont {Pedesseau},
  \citenamefont {Even}, \citenamefont {Chimiques}, \citenamefont {Uab},\ and\
  \citenamefont {Umr}}]{Robles2015a}%
  \BibitemOpen
  \bibfield  {author} {\bibinfo {author} {\bibfnamefont {R.}~\bibnamefont
  {Robles}}, \bibinfo {author} {\bibfnamefont {C.}~\bibnamefont {Katan}},
  \bibinfo {author} {\bibfnamefont {D.}~\bibnamefont {Sapori}}, \bibinfo
  {author} {\bibfnamefont {L.}~\bibnamefont {Pedesseau}}, \bibinfo {author}
  {\bibfnamefont {J.}~\bibnamefont {Even}}, \bibinfo {author} {\bibfnamefont
  {S.}~\bibnamefont {Chimiques}}, \bibinfo {author} {\bibfnamefont
  {C.}~\bibnamefont {Uab}}, \ and\ \bibinfo {author} {\bibfnamefont
  {F.}~\bibnamefont {Umr}},\ }\href {\doibase 10.1021/acsnano.5b04409}
  {\bibfield  {journal} {\bibinfo  {journal} {ACS Nano}\ }\textbf {\bibinfo
  {volume} {9}},\ \bibinfo {pages} {11557} (\bibinfo {year}
  {2015})}\BibitemShut {NoStop}%
\bibitem [{\citenamefont {Zheng}\ \emph {et~al.}(2015)\citenamefont {Zheng},
  \citenamefont {Tan}, \citenamefont {Liu},\ and\ \citenamefont
  {Rappe}}]{Zheng2015}%
  \BibitemOpen
  \bibfield  {author} {\bibinfo {author} {\bibfnamefont {F.}~\bibnamefont
  {Zheng}}, \bibinfo {author} {\bibfnamefont {L.~Z.}\ \bibnamefont {Tan}},
  \bibinfo {author} {\bibfnamefont {S.}~\bibnamefont {Liu}}, \ and\ \bibinfo
  {author} {\bibfnamefont {A.~M.}\ \bibnamefont {Rappe}},\ }\href@noop {}
  {\bibfield  {journal} {\bibinfo  {journal} {Nano Lett.}\ }\textbf {\bibinfo
  {volume} {15}},\ \bibinfo {pages} {7794} (\bibinfo {year} {2015})}\BibitemShut {NoStop}%
\bibitem [{\citenamefont {Etienne}\ \emph {et~al.}(2016)\citenamefont
  {Etienne}, \citenamefont {Mosconi},\ and\ \citenamefont {{De
  Angelis}}}]{Etienne2016a}%
  \BibitemOpen
  \bibfield  {author} {\bibinfo {author} {\bibfnamefont {T.}~\bibnamefont
  {Etienne}}, \bibinfo {author} {\bibfnamefont {E.}~\bibnamefont {Mosconi}}, \
  and\ \bibinfo {author} {\bibfnamefont {F.}~\bibnamefont {{De Angelis}}},\
  }\href {http://dx.doi.org/10.1021/acs.jpclett.6b00564} {\bibfield  {journal}
  {\bibinfo  {journal} {J. Phys. Chem. Lett.}\ }\textbf {\bibinfo {volume}
  {7}},\ \bibinfo {pages} {1638} (\bibinfo {year} {2016})}\BibitemShut
  {NoStop}%
\bibitem [{\citenamefont {Liu}\ \emph {et~al.}(2016)\citenamefont {Liu},
  \citenamefont {Kim}, \citenamefont {Tan},\ and\ \citenamefont
  {Rappe}}]{Liu2016}%
  \BibitemOpen
  \bibfield  {author} {\bibinfo {author} {\bibfnamefont {S.}~\bibnamefont
  {Liu}}, \bibinfo {author} {\bibfnamefont {Y.}~\bibnamefont {Kim}}, \bibinfo
  {author} {\bibfnamefont {L.~Z.}\ \bibnamefont {Tan}}, \ and\ \bibinfo
  {author} {\bibfnamefont {A.~M.}\ \bibnamefont {Rappe}},\ }\href {\doibase
  10.1021/acs.nanolett.5b04545} {\bibfield  {journal} {\bibinfo  {journal}
  {Nano Lett.}\ }\textbf {\bibinfo {volume} {16}},\ \bibinfo {pages} {1663}
  (\bibinfo {year} {2016})}\BibitemShut {NoStop}%
\bibitem [{\citenamefont {Dresselhaus}(1955)}]{Dresselhaus1955}%
  \BibitemOpen
  \bibfield  {author} {\bibinfo {author} {\bibfnamefont {G.}~\bibnamefont
  {Dresselhaus}},\ }\href@noop {} {\bibfield  {journal} {\bibinfo  {journal}
  {Phys. Rev. B}\ }\textbf {\bibinfo {volume} {100}},\ \bibinfo {pages} {580}
  (\bibinfo {year} {1955})}\BibitemShut {NoStop}%
\bibitem [{\citenamefont {Rashba}(1960)}]{Rashba1960}%
  \BibitemOpen
  \bibfield  {author} {\bibinfo {author} {\bibfnamefont {E.}~\bibnamefont
  {Rashba}},\ }\href@noop {} {\bibfield  {journal} {\bibinfo  {journal} {Sov.
  Phys. Solid State}\ }\textbf {\bibinfo {volume} {2}},\ \bibinfo {pages}
  {1109} (\bibinfo {year} {1960})}\BibitemShut {NoStop}%
\bibitem [{\citenamefont {Manchon}\ \emph {et~al.}(2015)\citenamefont
  {Manchon}, \citenamefont {Koo}, \citenamefont {Nitta}, \citenamefont
  {Frolov},\ and\ \citenamefont {Duine}}]{Manchon2015}%
  \BibitemOpen
  \bibfield  {author} {\bibinfo {author} {\bibfnamefont {A.}~\bibnamefont
  {Manchon}}, \bibinfo {author} {\bibfnamefont {H.~C.}\ \bibnamefont {Koo}},
  \bibinfo {author} {\bibfnamefont {J.}~\bibnamefont {Nitta}}, \bibinfo
  {author} {\bibfnamefont {S.~M.}\ \bibnamefont {Frolov}}, \ and\ \bibinfo
  {author} {\bibfnamefont {R.~A.}\ \bibnamefont {Duine}},\ }\href {\doibase
  10.1038/nmat4360} {\bibfield  {journal} {\bibinfo  {journal} {Nat. Mat.}\
  }\textbf {\bibinfo {volume} {14}},\ \bibinfo {pages} {871} (\bibinfo {year}
  {2015})}\BibitemShut {NoStop}%
\bibitem [{\citenamefont {Ishizaka}\ \emph {et~al.}(2011)\citenamefont
  {Ishizaka}, \citenamefont {Bahramy}, \citenamefont {Murakawa}, \citenamefont
  {Sakano}, \citenamefont {Shimojima}, \citenamefont {Sonobe}, \citenamefont
  {Koizumi}, \citenamefont {Shin}, \citenamefont {Miyahara}, \citenamefont
  {Kimura}, \citenamefont {Miyamoto}, \citenamefont {Okuda}, \citenamefont
  {Namatame}, \citenamefont {Taniguchi}, \citenamefont {Arita}, \citenamefont
  {Nagaosa}, \citenamefont {Kobayashi}, \citenamefont {Murakami}, \citenamefont
  {Kumai}, \citenamefont {Kaneko}, \citenamefont {Onose},\ and\ \citenamefont
  {Tokura}}]{Ishizaka2011}%
  \BibitemOpen
  \bibfield  {author} {\bibinfo {author} {\bibfnamefont {K.}~\bibnamefont
  {Ishizaka}}, \bibinfo {author} {\bibfnamefont {M.~S.}\ \bibnamefont
  {Bahramy}}, \bibinfo {author} {\bibfnamefont {H.}~\bibnamefont {Murakawa}},
  \bibinfo {author} {\bibfnamefont {M.}~\bibnamefont {Sakano}}, \bibinfo
  {author} {\bibfnamefont {T.}~\bibnamefont {Shimojima}}, \bibinfo {author}
  {\bibfnamefont {T.}~\bibnamefont {Sonobe}}, \bibinfo {author} {\bibfnamefont
  {K.}~\bibnamefont {Koizumi}}, \bibinfo {author} {\bibfnamefont
  {S.}~\bibnamefont {Shin}}, \bibinfo {author} {\bibfnamefont {H.}~\bibnamefont
  {Miyahara}}, \bibinfo {author} {\bibfnamefont {A.}~\bibnamefont {Kimura}},
  \bibinfo {author} {\bibfnamefont {K.}~\bibnamefont {Miyamoto}}, \bibinfo
  {author} {\bibfnamefont {T.}~\bibnamefont {Okuda}}, \bibinfo {author}
  {\bibfnamefont {H.}~\bibnamefont {Namatame}}, \bibinfo {author}
  {\bibfnamefont {M.}~\bibnamefont {Taniguchi}}, \bibinfo {author}
  {\bibfnamefont {R.}~\bibnamefont {Arita}}, \bibinfo {author} {\bibfnamefont
  {N.}~\bibnamefont {Nagaosa}}, \bibinfo {author} {\bibfnamefont
  {K.}~\bibnamefont {Kobayashi}}, \bibinfo {author} {\bibfnamefont
  {Y.}~\bibnamefont {Murakami}}, \bibinfo {author} {\bibfnamefont
  {R.}~\bibnamefont {Kumai}}, \bibinfo {author} {\bibfnamefont
  {Y.}~\bibnamefont {Kaneko}}, \bibinfo {author} {\bibfnamefont
  {Y.}~\bibnamefont {Onose}}, \ and\ \bibinfo {author} {\bibfnamefont
  {Y.}~\bibnamefont {Tokura}},\ }\href {\doibase 10.1038/nmat3051} {\bibfield
  {journal} {\bibinfo  {journal} {Nat. Mater.}\ }\textbf {\bibinfo {volume}
  {10}},\ \bibinfo {pages} {521} (\bibinfo {year} {2011})}\BibitemShut
  {NoStop}%
\bibitem [{\citenamefont {Kim}\ \emph {et~al.}(2014)\citenamefont {Kim},
  \citenamefont {Im}, \citenamefont {Freeman}, \citenamefont {Ihm},\ and\
  \citenamefont {Jin}}]{Kim2014f}%
  \BibitemOpen
  \bibfield  {author} {\bibinfo {author} {\bibfnamefont {M.}~\bibnamefont
  {Kim}}, \bibinfo {author} {\bibfnamefont {J.}~\bibnamefont {Im}}, \bibinfo
  {author} {\bibfnamefont {a.~J.}\ \bibnamefont {Freeman}}, \bibinfo {author}
  {\bibfnamefont {J.}~\bibnamefont {Ihm}}, \ and\ \bibinfo {author}
  {\bibfnamefont {H.}~\bibnamefont {Jin}},\ }\href {\doibase
  10.1073/pnas.1405780111} {\bibfield  {journal} {\bibinfo  {journal} {Proc.
  Nat. Acad. Sci.}\ }\textbf {\bibinfo {volume} {111}},\ \bibinfo {pages}
  {6900} (\bibinfo {year} {2014})}\BibitemShut {NoStop}%
\bibitem [{\citenamefont {Amat}\ \emph {et~al.}(2014)\citenamefont {Amat},
  \citenamefont {Mosconi}, \citenamefont {Ronca}, \citenamefont {Quarti},
  \citenamefont {Umari}, \citenamefont {Nazeeruddin}, \citenamefont {Gratzel},\
  and\ \citenamefont {Angelis}}]{Amat2014}%
  \BibitemOpen
  \bibfield  {author} {\bibinfo {author} {\bibfnamefont {A.}~\bibnamefont
  {Amat}}, \bibinfo {author} {\bibfnamefont {E.}~\bibnamefont {Mosconi}},
  \bibinfo {author} {\bibfnamefont {E.}~\bibnamefont {Ronca}}, \bibinfo
  {author} {\bibfnamefont {C.}~\bibnamefont {Quarti}}, \bibinfo {author}
  {\bibfnamefont {P.}~\bibnamefont {Umari}}, \bibinfo {author} {\bibfnamefont
  {M.~K.}\ \bibnamefont {Nazeeruddin}}, \bibinfo {author} {\bibfnamefont
  {M.}~\bibnamefont {Gratzel}}, \ and\ \bibinfo {author} {\bibfnamefont
  {F.~D.}\ \bibnamefont {Angelis}},\ }\href {\doibase 10.1021/nl5012992}
  {\bibfield  {journal} {\bibinfo  {journal} {Nano Lett.}\ }\textbf {\bibinfo
  {volume} {14}},\ \bibinfo {pages} {3608} (\bibinfo {year}
  {2014})}\BibitemShut {NoStop}%
\bibitem [{\citenamefont {Beilsten-Edmands}\ \emph {et~al.}(2015)\citenamefont
  {Beilsten-Edmands}, \citenamefont {Eperon}, \citenamefont {Johnson},
  \citenamefont {Snaith},\ and\ \citenamefont
  {Radaelli}}]{Beilsten-Edmands2015}%
  \BibitemOpen
  \bibfield  {author} {\bibinfo {author} {\bibfnamefont {J.}~\bibnamefont
  {Beilsten-Edmands}}, \bibinfo {author} {\bibfnamefont {G.~E.}\ \bibnamefont
  {Eperon}}, \bibinfo {author} {\bibfnamefont {R.~D.}\ \bibnamefont {Johnson}},
  \bibinfo {author} {\bibfnamefont {H.~J.}\ \bibnamefont {Snaith}}, \ and\
  \bibinfo {author} {\bibfnamefont {P.~G.}\ \bibnamefont {Radaelli}},\ }\href
  {\doibase 10.1063/1.4919109} {\bibfield  {journal} {\bibinfo  {journal}
  {Appl. Phys. Lett.}\ }\textbf {\bibinfo {volume} {106}},\ \bibinfo {pages}
  {173502} (\bibinfo {year} {2015})}\BibitemShut {NoStop}%
\bibitem [{\citenamefont {Fan}\ \emph {et~al.}(2015)\citenamefont {Fan},
  \citenamefont {Xiao}, \citenamefont {Sun}, \citenamefont {Chen},
  \citenamefont {Hu}, \citenamefont {Ouyang}, \citenamefont {Ong},
  \citenamefont {Zeng},\ and\ \citenamefont {Wang}}]{Fan2015}%
  \BibitemOpen
  \bibfield  {author} {\bibinfo {author} {\bibfnamefont {Z.}~\bibnamefont
  {Fan}}, \bibinfo {author} {\bibfnamefont {J.}~\bibnamefont {Xiao}}, \bibinfo
  {author} {\bibfnamefont {K.}~\bibnamefont {Sun}}, \bibinfo {author}
  {\bibfnamefont {L.}~\bibnamefont {Chen}}, \bibinfo {author} {\bibfnamefont
  {Y.}~\bibnamefont {Hu}}, \bibinfo {author} {\bibfnamefont {J.}~\bibnamefont
  {Ouyang}}, \bibinfo {author} {\bibfnamefont {K.~P.}\ \bibnamefont {Ong}},
  \bibinfo {author} {\bibfnamefont {K.}~\bibnamefont {Zeng}}, \ and\ \bibinfo
  {author} {\bibfnamefont {J.}~\bibnamefont {Wang}},\ }\href {\doibase
  10.1021/acs.jpclett.5b00389} {\bibfield  {journal} {\bibinfo  {journal} {J.
  Chem. Phys. Lett.}\ }\textbf {\bibinfo {volume} {6}},\ \bibinfo {pages}
  {1155} (\bibinfo {year} {2015})}\BibitemShut {NoStop}%
\bibitem [{\citenamefont {Govinda}\ \emph {et~al.}(2016)\citenamefont
  {Govinda}, \citenamefont {Mahale}, \citenamefont {Kore}, \citenamefont
  {Mukherjee}, \citenamefont {Pavan}, \citenamefont {De}, \citenamefont
  {Ghara}, \citenamefont {Sundaresan}, \citenamefont {Pandey}, \citenamefont
  {{Guru Row}},\ and\ \citenamefont {Sarma}}]{Govinda2016}%
  \BibitemOpen
  \bibfield  {author} {\bibinfo {author} {\bibfnamefont {S.}~\bibnamefont
  {Govinda}}, \bibinfo {author} {\bibfnamefont {P.}~\bibnamefont {Mahale}},
  \bibinfo {author} {\bibfnamefont {B.~P.}\ \bibnamefont {Kore}}, \bibinfo
  {author} {\bibfnamefont {S.}~\bibnamefont {Mukherjee}}, \bibinfo {author}
  {\bibfnamefont {M.~S.}\ \bibnamefont {Pavan}}, \bibinfo {author}
  {\bibfnamefont {C.}~\bibnamefont {De}}, \bibinfo {author} {\bibfnamefont
  {S.}~\bibnamefont {Ghara}}, \bibinfo {author} {\bibfnamefont
  {A.}~\bibnamefont {Sundaresan}}, \bibinfo {author} {\bibfnamefont
  {A.}~\bibnamefont {Pandey}}, \bibinfo {author} {\bibfnamefont {T.~N.}\
  \bibnamefont {{Guru Row}}}, \ and\ \bibinfo {author} {\bibfnamefont {D.~D.}\
  \bibnamefont {Sarma}},\ }\href {\doibase 10.1021/acs.jpclett.6b00803}
  {\bibfield  {journal} {\bibinfo  {journal} {J. Phys. Chem. Lett.}\ }\textbf
  {\bibinfo {volume} {7}},\ \bibinfo {pages} {2412} (\bibinfo {year}
  {2016})}\BibitemShut {NoStop}%
\bibitem [{Hoq(2016)}]{Hoque2016a}%
  \BibitemOpen
  \href@noop {} {\bibfield  {journal} {\bibinfo  {journal} {ACS Energy Lett.}\
  }\textbf {\bibinfo {volume} {1}},\ \bibinfo {pages} {142} (\bibinfo {year}
  {2016})}\BibitemShut {NoStop}%
\bibitem [{\citenamefont {Filippetti}\ \emph {et~al.}(2015)\citenamefont
  {Filippetti}, \citenamefont {Delugas}, \citenamefont {Saba},\ and\
  \citenamefont {Mattoni}}]{Filippetti2015a}%
  \BibitemOpen
  \bibfield  {author} {\bibinfo {author} {\bibfnamefont {A.}~\bibnamefont
  {Filippetti}}, \bibinfo {author} {\bibfnamefont {P.}~\bibnamefont {Delugas}},
  \bibinfo {author} {\bibfnamefont {M.~I.}\ \bibnamefont {Saba}}, \ and\
  \bibinfo {author} {\bibfnamefont {A.}~\bibnamefont {Mattoni}},\ }\href
  {http://dx.doi.org/10.1021/acs.jpclett.5b02117} {\bibfield  {journal}
  {\bibinfo  {journal} {J. Phys. Chem. Lett.}\ }\textbf {\bibinfo {volume}
  {6}},\ \bibinfo {pages} {4909} (\bibinfo {year} {2015})}\BibitemShut
  {NoStop}%
\bibitem [{\citenamefont {Gesi}(1997)}]{Gesi1997}%
  \BibitemOpen
  \bibfield  {author} {\bibinfo {author} {\bibfnamefont {K.}~\bibnamefont
  {Gesi}},\ }\href {\doibase 10.1080/00150199708012851} {\bibfield  {journal}
  {\bibinfo  {journal} {Ferroelectrics}\ }\textbf {\bibinfo {volume} {203}},\
  \bibinfo {pages} {249} (\bibinfo {year} {1997})}\BibitemShut {NoStop}%
\bibitem [{\citenamefont {Stoumpos}\ \emph {et~al.}(2013)\citenamefont
  {Stoumpos}, \citenamefont {Malliakas},\ and\ \citenamefont
  {Kanatzidis}}]{Stoumpos2013a}%
  \BibitemOpen
  \bibfield  {author} {\bibinfo {author} {\bibfnamefont {C.~C.}\ \bibnamefont
  {Stoumpos}}, \bibinfo {author} {\bibfnamefont {C.~D.}\ \bibnamefont
  {Malliakas}}, \ and\ \bibinfo {author} {\bibfnamefont {M.~G.}\ \bibnamefont
  {Kanatzidis}},\ }\href {\doibase 10.1021/ic401215x} {\bibfield  {journal}
  {\bibinfo  {journal} {Inorg. Chem.}\ }\textbf {\bibinfo {volume} {52}},\
  \bibinfo {pages} {9019} (\bibinfo {year} {2013})}\BibitemShut {NoStop}%
\bibitem [{\citenamefont {Kutes}\ \emph {et~al.}(2014)\citenamefont {Kutes},
  \citenamefont {Ye}, \citenamefont {Zhou}, \citenamefont {Pang}, \citenamefont
  {Huey},\ and\ \citenamefont {Padture}}]{Kutes2014a}%
  \BibitemOpen
  \bibfield  {author} {\bibinfo {author} {\bibfnamefont {Y.}~\bibnamefont
  {Kutes}}, \bibinfo {author} {\bibfnamefont {L.}~\bibnamefont {Ye}}, \bibinfo
  {author} {\bibfnamefont {Y.}~\bibnamefont {Zhou}}, \bibinfo {author}
  {\bibfnamefont {S.}~\bibnamefont {Pang}}, \bibinfo {author} {\bibfnamefont
  {B.~D.}\ \bibnamefont {Huey}}, \ and\ \bibinfo {author} {\bibfnamefont
  {N.~P.}\ \bibnamefont {Padture}},\ }\href
  {http://pubs.acs.org/doi/abs/10.1021/jz501697b} {\bibfield  {journal}
  {\bibinfo  {journal} {J. Phys. Chem. Lett.}\ }\textbf {\bibinfo {volume}
  {5}},\ \bibinfo {pages} {3335} (\bibinfo {year} {2014})}\BibitemShut
  {NoStop}%
\bibitem [{\citenamefont {Poglitsch}\ and\ \citenamefont
  {Weber}(1987)}]{Poglitsch1987}%
  \BibitemOpen
  \bibfield  {author} {\bibinfo {author} {\bibfnamefont {A.}~\bibnamefont
  {Poglitsch}}\ and\ \bibinfo {author} {\bibfnamefont {D.}~\bibnamefont
  {Weber}},\ }\href {\doibase 10.1063/1.453467} {\bibfield  {journal} {\bibinfo
   {journal} {J. Chem. Phys.}\ }\textbf {\bibinfo {volume} {87}},\ \bibinfo
  {pages} {6373} (\bibinfo {year} {1987})}\BibitemShut {NoStop}%
\bibitem [{\citenamefont {Chen}\ \emph {et~al.}(2015)\citenamefont {Chen},
  \citenamefont {Foley}, \citenamefont {Ipek}, \citenamefont {Tyagi},
  \citenamefont {Copley}, \citenamefont {Brown}, \citenamefont {Choi},\ and\
  \citenamefont {Lee}}]{Chen2015q}%
  \BibitemOpen
  \bibfield  {author} {\bibinfo {author} {\bibfnamefont {T.}~\bibnamefont
  {Chen}}, \bibinfo {author} {\bibfnamefont {B.~J.}\ \bibnamefont {Foley}},
  \bibinfo {author} {\bibfnamefont {B.}~\bibnamefont {Ipek}}, \bibinfo {author}
  {\bibfnamefont {M.}~\bibnamefont {Tyagi}}, \bibinfo {author} {\bibfnamefont
  {J.~R.~D.}\ \bibnamefont {Copley}}, \bibinfo {author} {\bibfnamefont {C.~M.}\
  \bibnamefont {Brown}}, \bibinfo {author} {\bibfnamefont {J.~J.}\ \bibnamefont
  {Choi}}, \ and\ \bibinfo {author} {\bibfnamefont {S.~H.}\ \bibnamefont
  {Lee}},\ }\href {\doibase 10.1039/C5CP05348J} {\bibfield  {journal} {\bibinfo
   {journal} {Phys. Chem. Chem. Phys.}\ }\textbf {\bibinfo {volume} {17}},\
  \bibinfo {pages} {31278} (\bibinfo {year} {2015})}\BibitemShut {NoStop}%
\bibitem [{\citenamefont {Mattoni}\ \emph {et~al.}(2015)\citenamefont
  {Mattoni}, \citenamefont {Filippetti}, \citenamefont {Saba},\ and\
  \citenamefont {Delugas}}]{Mattoni2015c}%
  \BibitemOpen
  \bibfield  {author} {\bibinfo {author} {\bibfnamefont {A.}~\bibnamefont
  {Mattoni}}, \bibinfo {author} {\bibfnamefont {A.}~\bibnamefont {Filippetti}},
  \bibinfo {author} {\bibfnamefont {M.~I.}\ \bibnamefont {Saba}}, \ and\
  \bibinfo {author} {\bibfnamefont {P.}~\bibnamefont {Delugas}},\ }\href
  {\doibase 10.1021/acs.jpcc.5b04283} {\bibfield  {journal} {\bibinfo
  {journal} {J. Phys. Chem. C}\ }\textbf {\bibinfo {volume} {119}},\ \bibinfo
  {pages} {17421} (\bibinfo {year} {2015})}\BibitemShut {NoStop}%
\bibitem [{\citenamefont {Kawamura}\ \emph {et~al.}(2002)\citenamefont
  {Kawamura}, \citenamefont {Mashiyama},\ and\ \citenamefont
  {Hasebe}}]{Kawamura2002}%
  \BibitemOpen
  \bibfield  {author} {\bibinfo {author} {\bibfnamefont {Y.}~\bibnamefont
  {Kawamura}}, \bibinfo {author} {\bibfnamefont {H.}~\bibnamefont {Mashiyama}},
  \ and\ \bibinfo {author} {\bibfnamefont {K.}~\bibnamefont {Hasebe}},\ }\href
  {\doibase 10.1143/JPSJ.71.1694} {\bibfield  {journal} {\bibinfo  {journal}
  {J. Phys. Soc. Jap.}\ }\textbf {\bibinfo {volume} {71}},\ \bibinfo {pages}
  {1694} (\bibinfo {year} {2002})}\BibitemShut {NoStop}%
\bibitem [{\citenamefont {Lee}\ \emph {et~al.}(2015)\citenamefont {Lee},
  \citenamefont {Bristowe}, \citenamefont {Bristowe},\ and\ \citenamefont
  {Cheetham}}]{Lee2015}%
  \BibitemOpen
  \bibfield  {author} {\bibinfo {author} {\bibfnamefont {J.-H.}\ \bibnamefont
  {Lee}}, \bibinfo {author} {\bibfnamefont {N.}~\bibnamefont {Bristowe}},
  \bibinfo {author} {\bibfnamefont {P.}~\bibnamefont {Bristowe}}, \ and\
  \bibinfo {author} {\bibfnamefont {T.}~\bibnamefont {Cheetham}},\ }\href
  {http://pubs.rsc.org/en/content/articlehtml/2015/cc/c5cc00979k} {\bibfield
  {journal} {\bibinfo  {journal} {Chem. Commun.}\ } (\bibinfo {year}
  {2015})}\BibitemShut {NoStop}%
\bibitem [{\citenamefont {Quarti}\ \emph {et~al.}(2014)\citenamefont {Quarti},
  \citenamefont {Mosconi},\ and\ \citenamefont {{De Angelis}}}]{Quarti2014a}%
  \BibitemOpen
  \bibfield  {author} {\bibinfo {author} {\bibfnamefont {C.}~\bibnamefont
  {Quarti}}, \bibinfo {author} {\bibfnamefont {E.}~\bibnamefont {Mosconi}}, \
  and\ \bibinfo {author} {\bibfnamefont {F.}~\bibnamefont {{De Angelis}}},\
  }\href {http://dx.doi.org/10.1021/cm5032046} {\bibfield  {journal} {\bibinfo
  {journal} {Chem. Mater.}\ }\textbf {\bibinfo {volume} {26}},\ \bibinfo
  {pages} {6557} (\bibinfo {year} {2014})}\BibitemShut {NoStop}%
\bibitem [{\citenamefont {Bl\"{o}chl}(1994)}]{Blochl1994}%
  \BibitemOpen
  \bibfield  {author} {\bibinfo {author} {\bibfnamefont {P.~E.}\ \bibnamefont
  {Bl\"{o}chl}},\ }\href@noop {} {\bibfield  {journal} {\bibinfo  {journal}
  {Phys. Rev. B}\ }\textbf {\bibinfo {volume} {50}},\ \bibinfo {pages} {17953}
  (\bibinfo {year} {1994})}\BibitemShut {NoStop}%
\bibitem [{\citenamefont {Kresse}\ and\ \citenamefont
  {Joubert}(1999)}]{Kresse1999}%
  \BibitemOpen
  \bibfield  {author} {\bibinfo {author} {\bibfnamefont {G.}~\bibnamefont
  {Kresse}}\ and\ \bibinfo {author} {\bibfnamefont {D.}~\bibnamefont
  {Joubert}},\ }\href {\doibase 10.1103/PhysRevB.59.1758} {\bibfield  {journal}
  {\bibinfo  {journal} {Phys. Rev. B}\ }\textbf {\bibinfo {volume} {59}},\
  \bibinfo {pages} {1758} (\bibinfo {year} {1999})}\BibitemShut {NoStop}%
\bibitem [{\citenamefont {Kresse}\ and\ \citenamefont
  {Hafner}(1993)}]{Kresse1993}%
  \BibitemOpen
  \bibfield  {author} {\bibinfo {author} {\bibfnamefont {G.}~\bibnamefont
  {Kresse}}\ and\ \bibinfo {author} {\bibfnamefont {J.}~\bibnamefont
  {Hafner}},\ }\href {\doibase 10.1103/PhysRevB.47.558} {\bibfield  {journal}
  {\bibinfo  {journal} {Phys. Rev. B}\ }\textbf {\bibinfo {volume} {47}},\
  \bibinfo {pages} {558} (\bibinfo {year} {1993})}\BibitemShut {NoStop}%
\bibitem [{\citenamefont {Kresse}\ and\ \citenamefont
  {Furthm\"{u}ller}(1996)}]{Kresse1996}%
  \BibitemOpen
  \bibfield  {author} {\bibinfo {author} {\bibfnamefont {G.}~\bibnamefont
  {Kresse}}\ and\ \bibinfo {author} {\bibfnamefont {J.}~\bibnamefont
  {Furthm\"{u}ller}},\ }\href {http://www.ncbi.nlm.nih.gov/pubmed/9984901}
  {\bibfield  {journal} {\bibinfo  {journal} {Phys. Rev. B}\ }\textbf {\bibinfo
  {volume} {54}},\ \bibinfo {pages} {11169} (\bibinfo {year}
  {1996})}\BibitemShut {NoStop}%
\bibitem [{\citenamefont {Els\"{a}sser}\ \emph {et~al.}(1994)\citenamefont
  {Els\"{a}sser}, \citenamefont {F\"{a}hnle}, \citenamefont {Chan},\ and\
  \citenamefont {Ho}}]{Elsasser1994}%
  \BibitemOpen
  \bibfield  {author} {\bibinfo {author} {\bibfnamefont {C.}~\bibnamefont
  {Els\"{a}sser}}, \bibinfo {author} {\bibfnamefont {M.}~\bibnamefont
  {F\"{a}hnle}}, \bibinfo {author} {\bibfnamefont {C.~T.}\ \bibnamefont
  {Chan}}, \ and\ \bibinfo {author} {\bibfnamefont {K.~M.}\ \bibnamefont
  {Ho}},\ }\href {\doibase 10.1103/PhysRevB.49.13975} {\bibfield  {journal}
  {\bibinfo  {journal} {Phys. Rev. B}\ }\textbf {\bibinfo {volume} {49}},\
  \bibinfo {pages} {13975} (\bibinfo {year} {1994})}\BibitemShut {NoStop}%
\bibitem [{\citenamefont {Egger}\ and\ \citenamefont
  {Kronik}(2014)}]{Egger2014}%
  \BibitemOpen
  \bibfield  {author} {\bibinfo {author} {\bibfnamefont {D.~A.}\ \bibnamefont
  {Egger}}\ and\ \bibinfo {author} {\bibfnamefont {L.}~\bibnamefont {Kronik}},\
  }\href@noop {} {\bibfield  {journal} {\bibinfo  {journal} {J. Phys. Chem.
  Lett.}\ }\textbf {\bibinfo {volume} {5}},\ \bibinfo {pages} {2728} (\bibinfo
  {year} {2014})}\BibitemShut {NoStop}%
\bibitem [{\citenamefont {Tkatchenko}\ and\ \citenamefont
  {Scheffler}(2009)}]{Tkatchenko2009}%
  \BibitemOpen
  \bibfield  {author} {\bibinfo {author} {\bibfnamefont {A.}~\bibnamefont
  {Tkatchenko}}\ and\ \bibinfo {author} {\bibfnamefont {M.}~\bibnamefont
  {Scheffler}},\ }\href {\doibase 10.1103/PhysRevLett.102.073005} {\bibfield
  {journal} {\bibinfo  {journal} {Phys. Rev. Lett.}\ }\textbf {\bibinfo
  {volume} {102}},\ \bibinfo {pages} {073005} (\bibinfo {year}
  {2009})}\BibitemShut {NoStop}%
\bibitem [{\citenamefont {Li}\ \emph {et~al.}(2015)\citenamefont {Li},
  \citenamefont {Cooper}, \citenamefont {Giannini}, \citenamefont {Liu},
  \citenamefont {Toma},\ and\ \citenamefont {Sharp}}]{Li2015}%
  \BibitemOpen
  \bibfield  {author} {\bibinfo {author} {\bibfnamefont {Y.}~\bibnamefont
  {Li}}, \bibinfo {author} {\bibfnamefont {J.~K.}\ \bibnamefont {Cooper}},
  \bibinfo {author} {\bibfnamefont {C.}~\bibnamefont {Giannini}}, \bibinfo
  {author} {\bibfnamefont {Y.}~\bibnamefont {Liu}}, \bibinfo {author}
  {\bibfnamefont {F.~M.}\ \bibnamefont {Toma}}, \ and\ \bibinfo {author}
  {\bibfnamefont {I.~D.}\ \bibnamefont {Sharp}},\ }\href {\doibase
  10.1021/jz502720a} {\bibfield  {journal} {\bibinfo  {journal} {J. Phys. Chem.
  Lett.}\ }\textbf {\bibinfo {volume} {6}},\ \bibinfo {pages} {493} (\bibinfo
  {year} {2015})}\BibitemShut {NoStop}%
\bibitem [{\citenamefont {King-Smith}\ and\ \citenamefont
  {Vanderbilt}(1993)}]{King-Smith1993}%
  \BibitemOpen
  \bibfield  {author} {\bibinfo {author} {\bibfnamefont {R.~D.}\ \bibnamefont
  {King-Smith}}\ and\ \bibinfo {author} {\bibfnamefont {D.}~\bibnamefont
  {Vanderbilt}},\ }\href {\doibase 10.1103/PhysRevB.47.1651} {\bibfield
  {journal} {\bibinfo  {journal} {Phys. Rev. B}\ }\textbf {\bibinfo {volume}
  {47}},\ \bibinfo {pages} {1651} (\bibinfo {year} {1993})}\BibitemShut
  {NoStop}%
\bibitem [{\citenamefont {Lee}\ \emph {et~al.}(2016)\citenamefont {Lee},
  \citenamefont {Lee}, \citenamefont {Kong},\ and\ \citenamefont
  {Jang}}]{Lee2016c}%
  \BibitemOpen
  \bibfield  {author} {\bibinfo {author} {\bibfnamefont {J.~H.}\ \bibnamefont
  {Lee}}, \bibinfo {author} {\bibfnamefont {J.-H.}\ \bibnamefont {Lee}},
  \bibinfo {author} {\bibfnamefont {E.-H.}\ \bibnamefont {Kong}}, \ and\
  \bibinfo {author} {\bibfnamefont {H.~M.}\ \bibnamefont {Jang}},\ }\href
  {\doibase 10.1038/srep21687} {\bibfield  {journal} {\bibinfo  {journal} {Sci.
  Rep.}\ }\textbf {\bibinfo {volume} {6}},\ \bibinfo {pages} {21687} (\bibinfo
  {year} {2016})}\BibitemShut {NoStop}%
\bibitem [{\citenamefont {Leblebici}\ \emph {et~al.}(2016)\citenamefont
  {Leblebici}, \citenamefont {Leppert}, \citenamefont {Li}, \citenamefont
  {Reyes-Lillo}, \citenamefont {Wickenburg}, \citenamefont {Wong},
  \citenamefont {Lee}, \citenamefont {Melli}, \citenamefont {Ziegler},
  \citenamefont {Angell}, \citenamefont {Ogletree}, \citenamefont {Ashby},
  \citenamefont {Toma}, \citenamefont {Neaton}, \citenamefont {Sharp},\ and\
  \citenamefont {Weber-Bargioni}}]{Leblebici2016}%
  \BibitemOpen
  \bibfield  {author} {\bibinfo {author} {\bibfnamefont {S.~Y.}\ \bibnamefont
  {Leblebici}}, \bibinfo {author} {\bibfnamefont {L.}~\bibnamefont {Leppert}},
  \bibinfo {author} {\bibfnamefont {Y.}~\bibnamefont {Li}}, \bibinfo {author}
  {\bibfnamefont {S.~E.}\ \bibnamefont {Reyes-Lillo}}, \bibinfo {author}
  {\bibfnamefont {S.}~\bibnamefont {Wickenburg}}, \bibinfo {author}
  {\bibfnamefont {E.}~\bibnamefont {Wong}}, \bibinfo {author} {\bibfnamefont
  {J.}~\bibnamefont {Lee}}, \bibinfo {author} {\bibfnamefont {M.}~\bibnamefont
  {Melli}}, \bibinfo {author} {\bibfnamefont {D.}~\bibnamefont {Ziegler}},
  \bibinfo {author} {\bibfnamefont {D.~K.}\ \bibnamefont {Angell}}, \bibinfo
  {author} {\bibfnamefont {D.~F.}\ \bibnamefont {Ogletree}}, \bibinfo {author}
  {\bibfnamefont {P.~D.}\ \bibnamefont {Ashby}}, \bibinfo {author}
  {\bibfnamefont {F.~M.}\ \bibnamefont {Toma}}, \bibinfo {author}
  {\bibfnamefont {J.~B.}\ \bibnamefont {Neaton}}, \bibinfo {author}
  {\bibfnamefont {I.~D.}\ \bibnamefont {Sharp}}, \ and\ \bibinfo {author}
  {\bibnamefont {Weber-Bargioni}},\ }\href@noop {} {\bibfield  {journal}
  {\bibinfo  {journal} {Nat. Energy}\ }\textbf {\bibinfo {volume} {1}},\
  \bibinfo {pages} {16093} (\bibinfo {year} {2016})}\BibitemShut {NoStop}%
\bibitem [{\citenamefont {Eperon}\ \emph {et~al.}(2014)\citenamefont {Eperon},
  \citenamefont {Stranks}, \citenamefont {Menelaou}, \citenamefont {Johnston},
  \citenamefont {Herz},\ and\ \citenamefont {Snaith}}]{Eperon2014a}%
  \BibitemOpen
  \bibfield  {author} {\bibinfo {author} {\bibfnamefont {G.~E.}\ \bibnamefont
  {Eperon}}, \bibinfo {author} {\bibfnamefont {S.~D.}\ \bibnamefont {Stranks}},
  \bibinfo {author} {\bibfnamefont {C.}~\bibnamefont {Menelaou}}, \bibinfo
  {author} {\bibfnamefont {M.~B.}\ \bibnamefont {Johnston}}, \bibinfo {author}
  {\bibfnamefont {L.~M.}\ \bibnamefont {Herz}}, \ and\ \bibinfo {author}
  {\bibfnamefont {H.~J.}\ \bibnamefont {Snaith}},\ }\href {\doibase
  10.1039/c3ee43822h} {\bibfield  {journal} {\bibinfo  {journal} {Energy
  Environ. Sci.}\ }\textbf {\bibinfo {volume} {7}},\ \bibinfo {pages} {982}
  (\bibinfo {year} {2014})}\BibitemShut {NoStop}%
\bibitem [{\citenamefont {Rehman}\ \emph {et~al.}(2015)\citenamefont {Rehman},
  \citenamefont {Milot}, \citenamefont {Eperon}, \citenamefont {Wehrenfennig},
  \citenamefont {Boland}, \citenamefont {Snaith}, \citenamefont {Johnston},\
  and\ \citenamefont {Herz}}]{Rehman2015a}%
  \BibitemOpen
  \bibfield  {author} {\bibinfo {author} {\bibfnamefont {W.}~\bibnamefont
  {Rehman}}, \bibinfo {author} {\bibfnamefont {R.~L.}\ \bibnamefont {Milot}},
  \bibinfo {author} {\bibfnamefont {G.~E.}\ \bibnamefont {Eperon}}, \bibinfo
  {author} {\bibfnamefont {C.}~\bibnamefont {Wehrenfennig}}, \bibinfo {author}
  {\bibfnamefont {J.~L.}\ \bibnamefont {Boland}}, \bibinfo {author}
  {\bibfnamefont {H.~J.}\ \bibnamefont {Snaith}}, \bibinfo {author}
  {\bibfnamefont {M.~B.}\ \bibnamefont {Johnston}}, \ and\ \bibinfo {author}
  {\bibfnamefont {L.~M.}\ \bibnamefont {Herz}},\ }\href
  {http://doi.wiley.com/10.1002/adma.201502969} {\bibfield  {journal} {\bibinfo
   {journal} {Adv. Mater.}\ }\textbf {\bibinfo {volume} {27}},\ \bibinfo
  {pages} {7938} (\bibinfo {year} {2015})}\BibitemShut {NoStop}%
\bibitem [{\citenamefont {Kulkarni}\ \emph {et~al.}(2014)\citenamefont
  {Kulkarni}, \citenamefont {Baikie}, \citenamefont {Boix}, \citenamefont
  {Yantara}, \citenamefont {Mathews},\ and\ \citenamefont
  {Mhaisalkar}}]{Kulkarni2014a}%
  \BibitemOpen
  \bibfield  {author} {\bibinfo {author} {\bibfnamefont {S.~A.}\ \bibnamefont
  {Kulkarni}}, \bibinfo {author} {\bibfnamefont {T.}~\bibnamefont {Baikie}},
  \bibinfo {author} {\bibfnamefont {P.~P.}\ \bibnamefont {Boix}}, \bibinfo
  {author} {\bibfnamefont {N.}~\bibnamefont {Yantara}}, \bibinfo {author}
  {\bibfnamefont {N.}~\bibnamefont {Mathews}}, \ and\ \bibinfo {author}
  {\bibfnamefont {S.}~\bibnamefont {Mhaisalkar}},\ }\href {\doibase
  10.1039/c4ta00435c} {\bibfield  {journal} {\bibinfo  {journal} {J. Mater.
  Chem. A}\ }\textbf {\bibinfo {volume} {2}},\ \bibinfo {pages} {9221}
  (\bibinfo {year} {2014})}\BibitemShut {NoStop}%
\bibitem [{\citenamefont {Rabe}\ \emph {et~al.}(2007)\citenamefont {Rabe},
  \citenamefont {Dawber}, \citenamefont {Lichtensteiger}, \citenamefont {Ahn},\
  and\ \citenamefont {Triscone}}]{Rabe2007}%
  \BibitemOpen
  \bibfield  {author} {\bibinfo {author} {\bibfnamefont {K.~M.}\ \bibnamefont
  {Rabe}}, \bibinfo {author} {\bibfnamefont {M.}~\bibnamefont {Dawber}},
  \bibinfo {author} {\bibfnamefont {C.}~\bibnamefont {Lichtensteiger}},
  \bibinfo {author} {\bibfnamefont {C.~H.}\ \bibnamefont {Ahn}}, \ and\
  \bibinfo {author} {\bibfnamefont {J.-M.}\ \bibnamefont {Triscone}},\ }in\
  \href@noop {} {\emph {\bibinfo {booktitle} {Phys. Ferroelectr. - A Mod.
  Perspect.}}},\ \bibinfo {editor} {edited by\ \bibinfo {editor} {\bibfnamefont
  {K.~M.}\ \bibnamefont {Rabe}}, \bibinfo {editor} {\bibfnamefont {C.~H.}\
  \bibnamefont {Ahn}}, \ and\ \bibinfo {editor} {\bibfnamefont {J.-M.}\
  \bibnamefont {Triscone}}}\ (\bibinfo  {publisher} {Springer-Verlag},\
  \bibinfo {address} {Berlin, Heidelberg},\ \bibinfo {year} {2007})\ pp.\
  \bibinfo {pages} {10--11}\BibitemShut {NoStop}%
\bibitem [{\citenamefont {Grote}\ and\ \citenamefont
  {Berger}(2015)}]{Grote2015}%
  \BibitemOpen
  \bibfield  {author} {\bibinfo {author} {\bibfnamefont {C.}~\bibnamefont
  {Grote}}\ and\ \bibinfo {author} {\bibfnamefont {R.~F.}\ \bibnamefont
  {Berger}},\ }\href {\doibase 10.1021/acs.jpcc.5b07446} {\bibfield  {journal}
  {\bibinfo  {journal} {J. Phys. Chem. C}\ }\textbf {\bibinfo {volume} {119}},\
  \bibinfo {pages} {22832} (\bibinfo {year} {2015})}\BibitemShut {NoStop}%
\bibitem [{\citenamefont {Eperon}\ \emph {et~al.}(2015)\citenamefont {Eperon},
  \citenamefont {Paterno}, \citenamefont {Sutton}, \citenamefont {Zampetti},
  \citenamefont {Haghighirad}, \citenamefont {Cacialli},\ and\ \citenamefont
  {Snaith}}]{Eperon2015}%
  \BibitemOpen
  \bibfield  {author} {\bibinfo {author} {\bibfnamefont {G.~E.}\ \bibnamefont
  {Eperon}}, \bibinfo {author} {\bibfnamefont {G.~M.}\ \bibnamefont {Paterno}},
  \bibinfo {author} {\bibfnamefont {R.~J.}\ \bibnamefont {Sutton}}, \bibinfo
  {author} {\bibfnamefont {A.}~\bibnamefont {Zampetti}}, \bibinfo {author}
  {\bibfnamefont {A.~A.}\ \bibnamefont {Haghighirad}}, \bibinfo {author}
  {\bibfnamefont {F.}~\bibnamefont {Cacialli}}, \ and\ \bibinfo {author}
  {\bibfnamefont {H.~J.}\ \bibnamefont {Snaith}},\ }\href {\doibase
  10.1039/C5TA06398A} {\bibfield  {journal} {\bibinfo  {journal} {J. Mater.
  Chem. A}\ }\textbf {\bibinfo {volume} {3}},\ \bibinfo {pages} {19688}
  (\bibinfo {year} {2015})}\BibitemShut {NoStop}%
\bibitem [{\citenamefont {Pertsev}\ \emph {et~al.}(1998)\citenamefont
  {Pertsev}, \citenamefont {Zembilgotov},\ and\ \citenamefont
  {Tagantsev}}]{Pertsev1998}%
  \BibitemOpen
  \bibfield  {author} {\bibinfo {author} {\bibfnamefont {N.~A.}\ \bibnamefont
  {Pertsev}}, \bibinfo {author} {\bibfnamefont {A.~G.}\ \bibnamefont
  {Zembilgotov}}, \ and\ \bibinfo {author} {\bibfnamefont {A.~K.}\ \bibnamefont
  {Tagantsev}},\ }\href {\doibase 10.1103/PhysRevLett.80.1988} {\bibfield
  {journal} {\bibinfo  {journal} {Phys. Rev. Lett.}\ }\textbf {\bibinfo
  {volume} {80}},\ \bibinfo {pages} {1988} (\bibinfo {year}
  {1998})}\BibitemShut {NoStop}%
\bibitem [{\citenamefont {Di\'{e}guez}\ \emph {et~al.}(2005)\citenamefont
  {Di\'{e}guez}, \citenamefont {Rabe},\ and\ \citenamefont
  {Vanderbilt}}]{Dieguez2005}%
  \BibitemOpen
  \bibfield  {author} {\bibinfo {author} {\bibfnamefont {O.}~\bibnamefont
  {Di\'{e}guez}}, \bibinfo {author} {\bibfnamefont {K.~M.}\ \bibnamefont
  {Rabe}}, \ and\ \bibinfo {author} {\bibfnamefont {D.}~\bibnamefont
  {Vanderbilt}},\ }\href {\doibase 10.1103/PhysRevB.72.144101} {\bibfield
  {journal} {\bibinfo  {journal} {Phys. Rev. B}\ }\textbf {\bibinfo {volume}
  {72}},\ \bibinfo {pages} {144101} (\bibinfo {year} {2005})}\BibitemShut
  {NoStop}%
\bibitem [{\citenamefont {Protesescu}\ \emph {et~al.}(2015)\citenamefont
  {Protesescu}, \citenamefont {Yakunin}, \citenamefont {Bodnarchuk},
  \citenamefont {Krieg}, \citenamefont {Caputo}, \citenamefont {Hendon},
  \citenamefont {Yang}, \citenamefont {Walsh},\ and\ \citenamefont
  {Kovalenko}}]{Protesescu2015a}%
  \BibitemOpen
  \bibfield  {author} {\bibinfo {author} {\bibfnamefont {L.}~\bibnamefont
  {Protesescu}}, \bibinfo {author} {\bibfnamefont {S.}~\bibnamefont {Yakunin}},
  \bibinfo {author} {\bibfnamefont {M.~I.}\ \bibnamefont {Bodnarchuk}},
  \bibinfo {author} {\bibfnamefont {F.}~\bibnamefont {Krieg}}, \bibinfo
  {author} {\bibfnamefont {R.}~\bibnamefont {Caputo}}, \bibinfo {author}
  {\bibfnamefont {C.~H.}\ \bibnamefont {Hendon}}, \bibinfo {author}
  {\bibfnamefont {R.~X.}\ \bibnamefont {Yang}}, \bibinfo {author}
  {\bibfnamefont {A.}~\bibnamefont {Walsh}}, \ and\ \bibinfo {author}
  {\bibfnamefont {M.~V.}\ \bibnamefont {Kovalenko}},\ }\href {\doibase
  10.1021/nl5048779} {\bibfield  {journal} {\bibinfo  {journal} {Nano Lett.}\
  }\textbf {\bibinfo {volume} {15}},\ \bibinfo {pages} {3692} (\bibinfo {year}
  {2015})}\BibitemShut {NoStop}%
\bibitem [{\citenamefont {Reyes-Lillo}\ and\ \citenamefont
  {Rabe}(2013)}]{Reyes-Lillo2013}%
  \BibitemOpen
  \bibfield  {author} {\bibinfo {author} {\bibfnamefont {S.~E.}\ \bibnamefont
  {Reyes-Lillo}}\ and\ \bibinfo {author} {\bibfnamefont {K.~M.}\ \bibnamefont
  {Rabe}},\ }\href {\doibase 10.1103/PhysRevB.88.180102} {\bibfield  {journal}
  {\bibinfo  {journal} {Phys. Rev. B}\ }\textbf {\bibinfo {volume} {88}},\
  \bibinfo {pages} {180102(R)} (\bibinfo {year} {2013})}\BibitemShut {NoStop}%
\end{thebibliography}
\end{document}